\documentclass[%
reprint,
nofootinbib,
 amsmath,amssymb,
 aps,
]{revtex4-2}

\usepackage{pdflscape}

\usepackage[displaymath]{lineno}
\usepackage{graphicx}% Include figure files
\usepackage{dcolumn}% Align table columns on decimal point
\usepackage{bm}% bold math
\usepackage{hyperref}
\usepackage{siunitx}
\usepackage{changepage}
\usepackage{xcolor}
\usepackage{rotating}

\usepackage{etoolbox}
\usepackage{csquotes}

\makeatletter
\patchcmd{\frontmatter@abstract@produce}
  {\vskip200\p@\@plus1fil
   \penalty-200\relax
   \vskip-200\p@\@plus-1fil}
  {}
  {}
  {}
\makeatother

\usepackage{atbegshi,picture,tikz}
\AtBeginShipoutNext{\AtBeginShipoutUpperLeft{%
  \put(\dimexpr\paperwidth-1cm\relax,-1cm){\makebox[0pt][r]{Published in Phys. Rev. E \textbf{103}, 022408 (2021)}}}}%
\AtBeginShipoutNext{\AtBeginShipoutUpperLeft{%
  \put(\dimexpr\paperwidth-1cm\relax,-1.5cm){\makebox[0pt][r]{DOI: \href{https://doi.org/10.1103/PhysRevE.103.022408}{10.1103/PhysRevE.103.022408}}}}}%

\usepackage{pdfpages} % include pdfs
\usepackage{pgffor} % for loops

\makeatletter
\AtBeginDocument{\let\LS@rot\@undefined}
\makeatother

\def\supplementfilename{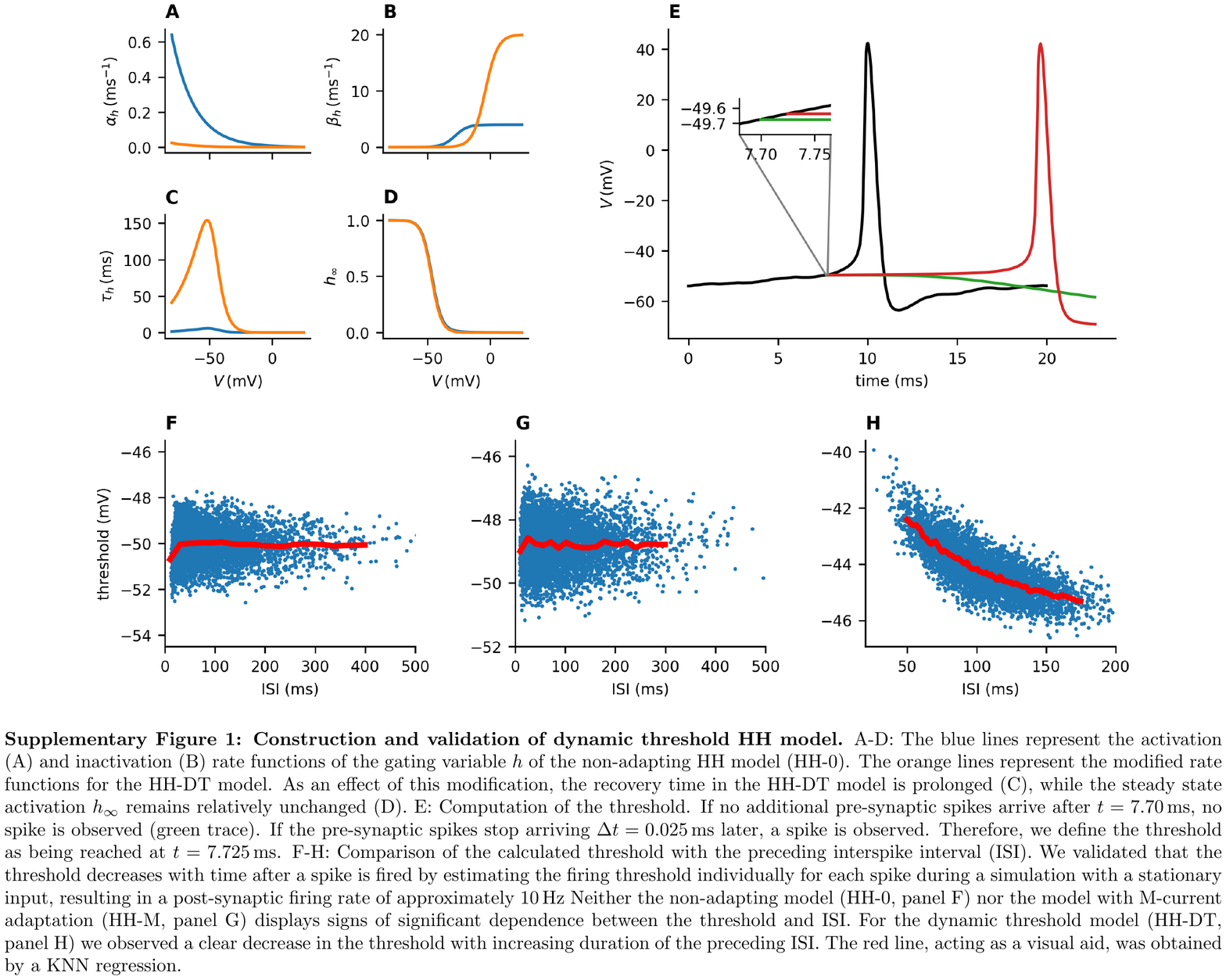}

\pdfximage{\supplementfilename}
\def\numbersupplementpages{\the\pdflastximagepages}

\newif\ifarXiv
\arXivtrue 
\begin{document}

% \preprint{APS/123-QED}
\title{Regular spiking in high conductance states: the essential role of inhibition}
\author{Tomas Barta}
\email{tomas.barta@fgu.cas.cz}
\affiliation{Institute of Physiology of the Czech Academy of Sciences, Prague, Czech Republic}
\affiliation{Charles University, First Medical Faculty,  Prague, Czech Republic}
\affiliation{Institute of Ecology and Environmental Sciences, INRA, Versailles, France}
\author{Lubomir Kostal}
\email{kostal@biomed.cas.cz}
\affiliation{Institute of Physiology of the Czech Academy of Sciences, Prague, Czech Republic}
%\date{\today}

\date{\today}% It is always \today, today,

\newcommand{\oder}[2]{\frac{\mathrm{d}#1}{\mathrm{d}#2}}
\newcommand{\exc}{\mathrm{e}}
\newcommand{\inh}{\mathrm{i}}
\newcommand{\musc}{\mathrm{M}}
\newcommand{\avgisi}{\mu_\mathrm{ISI}}
\newcommand{\avgisim}{\mu_\mathrm{ISI}^\AHP}
\newcommand{\stdisi}{\sigma_\mathrm{ISI}}
\newcommand{\cv}{\ifmmode \mathrm{C}_\mathrm{V}\else $\mathrm{C}_\mathrm{V}$\fi}
\newcommand{\gexc}{g_\exc}
\newcommand{\ginh}{g_\inh}
\newcommand{\gleak}{g_\mathrm{L}}
\newcommand{\gmusc}{g_\mathrm{M}}
\newcommand{\AHP}{\mathrm{AHP}}
\newcommand{\gahp}{g_\AHP}
\newcommand{\gtot}{g_\mathrm{tot}}
\newcommand{\gtotm}{g_\mathrm{tot}^\AHP}
\newcommand{\vef}{V_\mathrm{ef}}
\newcommand{\vefm}{V_\mathrm{ef}^\AHP}
\newcommand{\tef}{\tau_\mathrm{ef}}
\newcommand{\tefm}{\tau_\mathrm{ef}^\AHP}
\renewcommand{\K}{\mathrm{K}}
\newcommand{\Na}{\mathrm{Na}}
\newcommand{\EK}{E_\K}
\newcommand{\EL}{E_\mathrm{L}}
\newcommand{\Ee}{E_\mathrm{e}}
\newcommand{\Ei}{E_\mathrm{i}}
\newcommand{\limg}{\lim\limits_{\gexc^0\rightarrow+\infty}}
\newcommand{\supl}[1]{Supplementary Material, Section \ref{#1}}
\newcommand{\suplfigA}{Supplementary Figure 1}
\newcommand{\suplfigB}{Supplementary Figure 2}
\newcommand{\suplfigC}{Supplementary Figure 3}
\newcommand{\suplfigD}{Supplementary Figure 4}
\newcommand{\fluct}{\mathrm{F}}

\begin{abstract}
Strong inhibitory input to neurons, which occurs in balanced states of neural networks, increases synaptic current fluctuations. This has led to the assumption that inhibition contributes to the high spike-firing irregularity observed \textit{in vivo}. We used single compartment neuronal models with time-correlated (due to synaptic filtering) and state-dependent (due to reversal potentials) input to demonstrate that inhibitory input acts to decrease membrane potential fluctuations, a result that cannot be achieved with simplified neural input models. To clarify the effects on spike-firing regularity, we used models with different spike-firing adaptation mechanisms and observed that the addition of inhibition increased firing regularity in models with dynamic firing thresholds and decreased firing regularity if spike-firing adaptation was implemented through ionic currents or not at all. This novel fluctuation-stabilization mechanism provides a new perspective on the importance of strong inhibitory inputs observed in balanced states of neural networks and highlights the key roles of biologically plausible inputs and specific adaptation mechanisms in neuronal modeling.
\end{abstract}

\maketitle

\section*{Introduction}

%\copyrightnotice
In awake animals, neocortical neurons receive a stream of random synaptic inputs arising from background network activity \cite{Matsumura1988,Rudolph2007,Steriade2001}. This \textquote{synaptic noise} is responsible for the fluctuations in membrane potential and stochastic nature of spike-firing times \cite{Shadlen1994,Vreeswijk1996,Shadlen1998,Amit1997,Brunel2000,Destexhe2010,Machens2016}. Since spike-firing times encode the information transmitted by neurons, investigating the properties of neuronal responses to stochastic input, representing pre-synaptic spike arrivals, is of significant interest.

Typically, the total conductance of inhibitory synapses is several-fold higher than that of excitatory synapses \cite{Destexhe2003}. This state, commonly referred to as the \textquote{high conductance state} has been demonstrated to significantly affect the integrative properties of neurons \cite{Bernander1991,Pare1998,Mittmann2005,Wolfart2005,Rudolph2007}. Concurrently, the high inhibition-to-excitation ratio introduces additional synaptic noise, which should intuitively result in noisier firing. However, studies have demonstrated that the high ratio of inhibition may lead to more efficient information transmission \cite{Sengupta2013,DOnofrio2019,Barta2019}. \textit{In vivo} studies have also demonstrated that the onset of stimuli can stabilize the membrane potential without a significant change in its mean value \cite{Monier2003, Churchland2010}. Monier et al. \cite{Monier2003} observed that the decrease in fluctuations was associated with higher evoked inhibition, which may have a shunting effect \cite{Fatt1953}. Nevertheless, a theoretical framework explaining why and under which conditions this shunting effect overpowers the increased synaptic noise is lacking.
%\old{suggested that the suppression of fluctuations is due to shunting inhibitory input.}

Synaptic input can be modelled as temporary opening of excitatory and inhibitory ion channels, which act to either depolarize or hyperpolarize the neural membrane, respectively. Statistical measures of membrane potential can be calculated exactly with the resulting expressions being non-analytic \cite{Wolff2010} or they can be approximated in the steady-state with the effective time-constant approximation \cite{Richardson2004,Richardson2005}. For better analytical tractability, the synaptic drive is often simplified with one (or both) of the following assumptions:
\begin{enumerate}
    \item[A1] The magnitude of the synaptic current elicited by each presynaptic spike is independent of the voltage \cite{Lindner2001,Brunel2001,Fourcaud2002,Moreno-Bote2004,Schwalger2008,Droste2017}, or
    \item[A2] Time profiles of individual synapses (synaptic filtering) are neglected \cite{Brunel2000,Richardson2004,Richardson2006,Lanska1994,Deger2012,Droste2017,Sanzeni2020}.
\end{enumerate}

\begin{figure*}[htp]
    \centering
    \includegraphics[scale=1.]{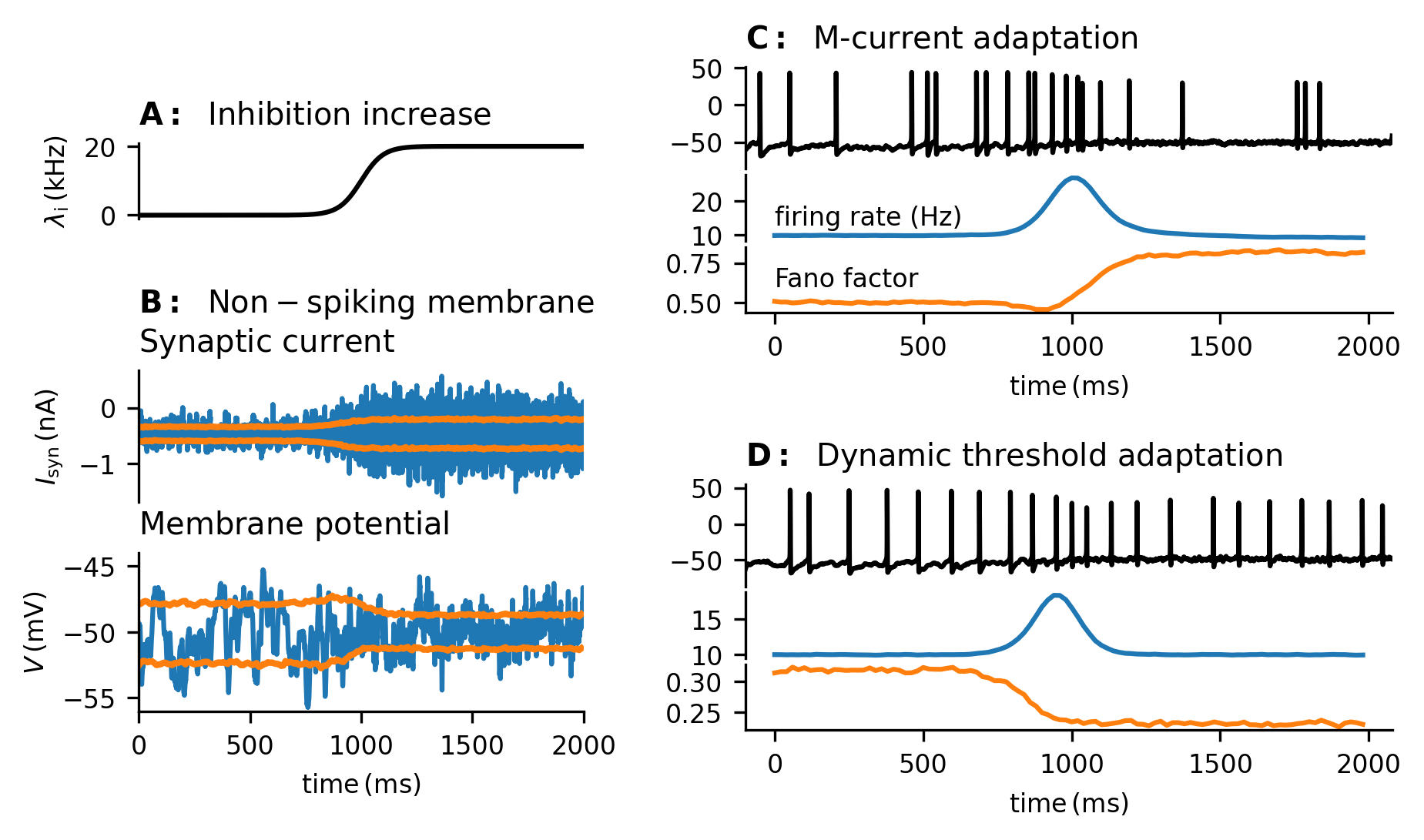}
    \caption{
    A: During a \SI{2}{\second} long simulation, the intensity of inhibitory input increases from \SI{0}{\kilo\hertz} to \SI{20}{\kilo\hertz}. The pre-synaptic spike trains are modeled as Poisson point processes. B: The intensity of excitatory input is increased simultaneously with the inhibition in order to maintain the mean membrane potential constant. This increases fluctuations of the synaptic current but decreases fluctuations of the membrane potential. The orange lines signify the mean value $\pm$ standard deviation. C and D: The effect of membrane potential stabilization on firing regularity. The intensity of the inhibitory input follows the time course shown in A, and the intensity of the excitatory activity is increased in order to maintain the steady state post-synaptic firing rate at approximately \SI{10}{\hertz} (blue trace). The firing regularity (measured here by the Fano factor, orange trace) decreases with the addition of inhibition to the input in the model with spike-firing adaptation by M currents (C). However, the model with dynamic threshold (D) exhibits a clear increase in regularity with the added inhibitory input. The mean firing rate and Fano factor were calculated by a sliding window of length \SI{100}{\milli\second} from approximately $10^5$ trials.}
    \label{fig:abstract}
\end{figure*}

In order to observe the shunting effect of inhibition \cite{Monier2003}, reversal potentials have to be considered, which excludes assumption A1. Richardson \cite{Richardson2004} demonstrated that an increase in inhibition could decrease the membrane potential for strongly hyperpolarized membranes in a model of synaptic input with omitted synaptic filtering (assumption A2). However, we demonstrate that if neither of the simplifying assumptions are used, the membrane potential stabilization effect can be observed across the complete range of membrane potentials, despite increased synaptic current fluctuations (Fig \ref{fig:abstract}A,B).

This naturally poses the question if the decreased membrane potential fluctuations lead to more regular firing activity \cite{Churchland2010}. To this end, we analyze the effect of membrane potential stabilization on different neuronal models. In particular, we focus on how the effects of inhibition change for different spike-firing adaptation (SFA) mechanisms. SFA is responsible for the decrease of a neuron's firing rate in response to a sustained stimulus and plays a crucial role in all stages of sensory processing (e.g., \citep{Martinez2005,Peron2009,Augustin2013,Ha2017,Levakova2019,Betkiewicz2020}). We compare two distinct SFA mechanisms: adaptation through ionic currents (muscarinic currents, AHP currents) and adaptation through dynamic threshold. We demonstrate that despite their formal similarities \citep{Benda2003,Kobayashi2016} the effect of inhibition qualitatively differs for these SFA mechanisms (Fig \ref{fig:abstract}C,D). We illustrate the differences on the analytically more tractable generalized leaky integrate-and-fire models (GLIF) followed by the biophysically more plausible Hodgkin-Huxley (HH)-type models.

\section*{Methods}
\label{s:matmet}

\subsection*{Subthreshold membrane potential}

In order to analyze the behavior of neurons in the absence of any spike-firing mechanism, we consider a point neuronal model with membrane potential $V$ described by

\begin{equation}
	C\oder{V(t)}{t}=-\gleak(V(t)-\EL)+\frac{1}{a}(I_\exc(t)+I_\inh(t)),
	\label{eq:mat_pot}
\end{equation}
where $C$ is the specific capacitance of the membrane, $\gleak$ is the specific leak conductance, $\EL$ is the leakage potential, $I_{\exc,\inh}$ are the synaptic currents due to stimulation by afferent neurons through excitatory and inhibitory synapses, respectively, and $a$ is the membrane area \cite{Tuckwell1988,DayanAbbot}. For brevity, we will further use $V\equiv V(t)$. The synaptic currents are described by
\begin{equation}
	I_{\exc,\inh}(t) = g_{\exc,\inh}(t)(V-E_{\exc,\inh}),
	\label{eq:isyn}
\end{equation}
where $\gexc(t)$, $\ginh(t)$ are the total excitatory and inhibitory conductances, and $\Ee$, $\Ei$ are the respective synaptic reversal potentials.

The total conductances in the Eq \eqref{eq:isyn} are given by
\begin{equation}
	g_{\exc,\inh}(t) = \sum_{t_k\in\mathcal{T}_{\exc,\inh}}h_{\exc,\inh}(t-t_k),
\end{equation}
where $\mathcal{T}_{\exc,\inh}$ are sets of presynaptic spike times modeled as realizations of stochastic point processes and $h_{\exc,\inh}$ are filtering functions (i.e., time profiles of individual excitatory and inhibitory conductances).

Unless stated otherwise, we used the following parameters: $C=\SI{1}{\micro\farad/\cm^2}$, $\gleak=\SI{0.045}{\milli\siemens/\cm^2}$, $\EL=\SI{-80}{\milli\volt}$, $\Ee=\SI{0}{\milli\volt}$, $\Ei=\SI{-75}{\milli\volt}$, $a=\SI{3.4636e-4}{\cm^2}$ \cite{Destexhe2001}.

\subsection*{Spike firing models}
\subsubsection*{GLIF models}
\label{s:params}

We consider three versions of the GLIF model:
\begin{enumerate}
    \item The classical Leaky Integrate-and-Fire model (LIF),
    \item LIF with SFA through ionic (after-hyperpolarization) currents (AHP-LIF),
    \item LIF with SFA through dynamic threshold (DT-LIF).
\end{enumerate}

The membrane potential of the LIF model obeys the Eq (\ref{eq:mat_pot}). Whenever $V>\theta$, where $\theta$ is a fixed threshold value, a spike is fired, and the membrane potential $V$ is reset to a value $V_r$. For our simulations, we used $\theta=\SI{-55}{\milli\volt}$ and $V_r=\EL$.

In the model with AHP current SFA (AHP-LIF), an additional hyperpolarizing conductance $\gahp$ is included in the model, and the membrane potential then obeys the equation \cite{Gerstner2019,Teeter2018,Benda2010}:
\begin{equation}
	C\oder{V}{t} = -\gleak(V-\EL) - \gahp(t)(V-\EK) + I_\exc(t) + I_\inh(t),
	\label{eq:mcurr}
\end{equation}
or equivalently
\begin{align}
    \tefm\oder{V}{t} &= -(V-\vefm), \\
    \tefm(t) &= \frac{aC}{\gexc(t)+\ginh(t)+\gahp(t)+a\gleak}, \label{eq:taum}\\
    \vefm(t) &= \frac{a\gleak\EL + \gexc(t)\Ee + \ginh(t)\Ei + \gahp(t)\EK}{\gexc(t)+\ginh(t)+\gahp(t)+a\gleak}.\label{eq:vefm}
\end{align}
where $\EK$ is the potassium reversal potential, and $\gahp$ is the corresponding conductance which increases by $\Delta \gahp$ when a spike is fired and otherwise decays exponentially to zero with a time constant $\tau_\AHP$. $\vefm$ then represents the effective reversal potential. Note that for simplicity, we omitted the voltage dependence of $\gahp$.

In the dynamic threshold model (DT-LIF), the threshold increases by $\Delta\theta$ after each spike and then decreases exponentially to $\theta_0$ with time constant $\tau_\theta$.

The parameters for the GLIF models are specified in the Tab. \ref{tab:glif_params}.

\begin{table}
    \centering
    \caption{{\bf GLIF models parameters}}
    \begin{tabular}{l|ccc}
                & LIF & AHP-LIF & DT-LIF \\\hline
        $\theta$, $\theta_0$\,(\si{\milli\volt})& -50 & -50 & -50 \\
        $\tau_\AHP$\,(\si{\milli\second}) & - & 100 & - \\
        $\Delta g_\AHP$\,(\si{\nano\siemens}) & 0 & 5 & 0 \\
        $\EK$\,(\si{\milli\volt}) &  - & -100 & - \\
        $\tau_\theta$\,(\si{\milli\second}) & - & - & 100 \\
        $\Delta_\theta$\,(\si{\milli\volt}) & 0 & 0 & 4
    \end{tabular}
    \label{tab:glif_params}
\end{table}

\subsubsection*{Hodgkin-Huxley models}
We adopted HH-type models developed by Destexhe et al. \cite{Destexhe1999}. The membrane potential obeys the equation:
\begin{equation}
\begin{split}
    C\oder{V}{t}=&-\gleak(V-\EL)-g_\Na m^3h(V-E_\Na)-\\
    &-g_\K n^4(V-E_\K)-g_\musc p(V-E_\K)-\frac{1}{a}I_\mathrm{syn},
\end{split}
\end{equation}
where $E_\Na$ and $E_\K$ are the sodium and potassium reversal potentials, respectively; $g_\Na$, $g_\K$, and $g_\musc$ are peak conductances; and $m$, $h$, $n$, and $p$ are gating variables obeying the equation:
\begin{equation}
    \oder{x}{t}=\alpha_x(V)(1-x) - \beta_x(V) x,
\end{equation}
or equivalently:
\begin{equation}
    \tau_x(V)\oder{x}{t}=-(x-x_\infty(V)),
\end{equation}
where $x$ is the respective gating variable, $\alpha_x$ and $\beta_x$ are the activation and inactivation functions, respectively, and
\begin{align}
    \tau_x(V) &= \frac{1}{\alpha_x(V)+\beta_x(V)}, \\
    x_\infty(V) &= \frac{\alpha_x(V)}{\alpha_x(V)+\beta_x(V)}.
\end{align}

The activation and inactivation functions are defined as follows:
\begin{align}
    \alpha_m &= -0.32 \frac{V-V_T-13}{\exp(-(V-V_T-13)/4)-1}, \\
    \beta_m &= 0.28 \frac{V-V_T-40}{\exp((V-V_T-40)/5)-1}, \\
    \alpha_h &= A_h\exp(-(V-V_T-V_S-17)/18), \\
    \beta_h &= \frac{4}{1+\exp(-(V-V_T-V_S-40)/5)}, \\
    \alpha_n &= -0.032 \frac{V - V_T - 15}{\exp(-(V - V_T - 15) / 5) - 1}, \\
    \beta_n &= 0.5  \exp(-(V - V_T - 10) / 40), \\
    \alpha_p &= 0.0001 \frac{ V + 30 } {1 - \exp(-(V + 30) / 9)}, \\
    \beta_p &= -0.0001 \frac{ V + 30 }  { 1 - \exp((V + 30) / 9) }.
\end{align}

\begin{table}[]
    \centering
    \caption{{\bf Parameters of the HH models}}
    \begin{tabular}{l|ccc}
                & HH-0 & HH-M & HH-DT \\\hline
        $g_\Na$\,(\si{\milli\siemens/\cm^2}) & 50 & 50 & 50 \\
        $g_\K$\,(\si{\milli\siemens/\cm^2}) &  5 &  5 &  5 \\
        $g_\musc$\,(\si{\milli\siemens/\cm^2}) &  0 &  0.5 &  0 \\
        $E_\Na$\,(\si{\milli\volt}) & 50 & 50 & 50 \\
        $\EK$\,(\si{\milli\volt}) & -90 & -90 & -90\\
        $V_T$\,(\si{\milli\volt}) & -58 & -58 & -58\\
        $V_S$\,(\si{\milli\volt}) & -10 & -10 & 14\\
        $A_h$\,(\si{\milli\second^{-1}}) & 0.128 & 0.128 & 0.00128
    \end{tabular}
    \label{tab:hh_params}
\end{table}

We set $g_M=0$ in both the HH-0 and HH-DT models, and $g_M>0$ in the HH-M model. In order to achieve dynamic threshold behavior, we modified the activation and deactivation functions of the gating variable $h$, which is responsible for deactivating voltage-gated sodium channels after firing a spike, by changing the parameters $A_h$ and $V_S$. For more details see the \suplfigA{}.

The parameters for the three HH-type models (without SFA (HH-0) / M-current SFA (HH-M) / dynamic threshold SFA (HH-DT)) are specified in the Tab \ref{tab:hh_params}.

\subsection*{Simulation details}
For synapses, we used the exponential filtering function:
\begin{equation}\label{eq:expon}
  h_{\exc,\inh}(t)=\begin{cases}
               A_{\exc,\inh}\exp(-t/\tau_{\exc,\inh})\ \ \ t\geq0\\
               0\ \ \ t<0
            \end{cases}
\end{equation}
with $A_\exc=A_\inh=\SI{0.0015}{\micro\siemens}$, $\tau_\exc=\SI{3}{\milli\second}$, $\tau_\exc=\SI{10}{\milli\second}$. Such input parameters with intensities $\lambda_\exc=\SI{2.67}{\hertz}$ and $\lambda_\inh=\SI{3.73}{\kilo\hertz}$ provide an input with $\gexc^0=\SI{12}{\nano\siemens}$, $\sigma_\exc=\SI{3}{\nano\siemens}$, $\ginh^0=\SI{57}{\nano\siemens}$, and $\sigma_\inh=\SI{6.6}{\nano\siemens}$, as reported by Destexhe et al. \cite{Destexhe2001}.

To ensure stability of the computation, we used the following update rule for the simulations:
\begin{align}
    V_{n+1} &= \left( \vef \right)_{n+1} + \left( V_n-\left( \vef \right)_{n+1} \right)\exp\left(\frac{\Delta t}{\tau_{n+1}}\right), \\
    \vef &= \frac{\sum_{x\in X}g_x E_x}{\sum_{x\in X}g_x} \\
    \tef &= \frac{C}{\sum_{x\in X}g_x}
\end{align}
where $X$ contains all the channel types (synaptic, leak, voltage-gated, and adaptive). The update rule for the synaptic conductances $g_{\exc,\inh}$ was
\begin{equation}
    \left( g_{\exc,\inh} \right)_{n+1} = \left( g_{\exc,\inh} \right)_{n}\exp\left(\frac{\Delta t}{\tau_{\exc,\inh}}\right)+ N_{\exc,\inh}A_{\exc,\inh},
\end{equation}
where $\left( N_{\exc,\inh} \right)$ is a Poisson random variable with mean $\lambda_{\exc,\inh}\Delta t$.

We used the step size $\Delta t=\SI{0.025}{\milli\second}$.

\subsection*{Evaluating firing rate regularity}

A classical measure of the firing regularity of steady spike trains is the coefficient of variation (\cv{}), defined as follows (e.g., \cite{Softky1993}):
\begin{equation}
	\cv{} = \frac{\sigma_{\mathrm{ISI}}}{\avgisi},\label{eq:cv}
\end{equation}
where $\avgisi$ and $\stdisi$ are the mean and standard deviation of the interspike intervals (ISIs), respectively. Lower \cv{} indicates higher firing regularity.

To achieve an accurate estimate of the \cv{}, we estimated the statistics from approximately 160,000 ISIs for each data point. For a Poisson process ($\cv{}=1$) with this number of ISIs, the estimate of $\cv{}$ falls within $[0.995,1.005]$ in over 95\% of cases. Note that the estimation was more accurate for lower values of $\cv{}$.

\section*{Results}

\subsection*{Membrane potential is stabilized with increased input fluctuations}

Since the inputs to a neuron consist of pooled spike trains from a large number of presynaptic neurons, according to the Palm-Khintchine theorem \cite{Heyman2004}, it is sufficient to approximate the excitatory and inhibitory inputs by Poisson processes with intensities $\lambda_\exc$ and $\lambda_\inh$, respectively \cite{Tuckwell1988}. It has been demonstrated that this condition is not necessarily satisfied for neurons \textit{in vivo} \cite{Lindner2006}. However, as we discuss below, this should not affect the conclusions of our analysis. According to Campbell's theorem \cite{Kingman1993}, it then holds for the mean $g_{\exc,\inh}^0$ and variance $\sigma^2_{\exc,\inh}$ of the input
\begin{align}
    g_{\exc,\inh}^0 &= \lambda_{\exc,\inh}\int_0^\infty h_{\exc,\inh}(t)\,\mathrm{d}t, \label{eq:cmpb1} \\
    \sigma_{\exc,\inh}^2 &= \lambda_{\exc,\inh}\int_0^\infty h_{\exc,\inh}^2(t)\,\mathrm{d}t. \label{eq:cmpb2}
\end{align}
Therefore $\frac{\sigma_{\exc,\inh}}{g_{\exc,\inh}^0}=O\left(\frac{1}{\sqrt{\lambda_{\exc,\inh}}}\right)$ (a well-known property of the Poisson shot noise \cite{Tuckwell1988})\linelabel{lne:wellknown}.

For the purposes of our analysis, we consider the voltage equations of a membrane without any spike-generating mechanism as:
\begin{align}
	\tef(\gexc(t),\ginh(t))\oder{V}{t} &= -V - \vef(\gexc(t),\ginh(t)), \label{eq:voltage} \\
	\tef(\gexc,\ginh) &= \frac{aC}{a\gleak+\gexc+\ginh}, \label{eq:tau}\\
	\vef(\gexc,\ginh) &= \frac{a\gleak \EL + \gexc \Ee+\ginh \Ei}{\gleak+\gexc+\ginh}. \label{eq:hatv}
\end{align}
For large inputs $\frac{\sigma_{\exc,\inh}}{g_{\exc,\inh}^0}\ll 1$, we can linearize the Eq (\ref{eq:hatv}):
\begin{equation}
    \vef(\gexc,\ginh) \doteq E_0 \left( 1-\frac{\gexc^\fluct+\ginh^\fluct}{a\gleak+\gexc^0+\ginh^0} \right) + \frac{\gexc^\fluct\Ee+\ginh^\fluct\Ei}{a\gleak+\gexc^0+\ginh^0}, \label{eq:veflin}
\end{equation}
where $E_0=\vef(\gexc^0,\ginh^0)$ and $g_{\exc,\inh}^\fluct=g_{\exc,\inh}-g_{\exc,\inh}^0$. Since the fluctuating terms in Eq (\ref{eq:veflin}) disappear with growing input, evaluating the limits with a fixed inhibition-to-excitation ratio $c=\frac{\ginh^0}{\gexc^0}$ leads to:
\begin{align}
	&\lim_{\lambda_\exc,\lambda_\inh\rightarrow\infty}\mathrm{E\left[\vef\right]}= V_\infty(c)\equiv\frac{E_\exc+cE_\inh}{1+c}, \label{eq:vinfty} \\
	&\lim_{\lambda_\exc,\lambda_\inh\rightarrow\infty}\mathrm{Var\left[\vef\right]}=0.
\end{align}
\linelabel{lne:upperbound_start}$\mathrm{Var\left[\vef\right]}$ is an upper bound on the variance of $V$ (it follows from the Eq \eqref{eq:voltage} that the membrane potential $V$ is essentially a \textquote{low-pass filtered} effective reversal potential $\vef$). Therefore, it also holds that
$\lim\limits_{\lambda_\exc,\lambda_\inh\rightarrow\infty}\mathrm{\langle V\rangle}= V_\infty(c)\equiv\frac{E_\exc+cE_\inh}{1+c}$ and $\lim\limits_{\lambda_\exc,\lambda_\inh\rightarrow\infty}\mathrm{\sigma_V}=0$\linelabel{lne:upperbound_end}. This can also be observed from the perturbative approach suggested in \cite{Amit1992} and further developed in \cite{Richardson2004,Richardson2005}. 
Therefore, any membrane potential between the reversal potentials $\Ei$, $\Ee$ can be asymptotically reached with zero variance, despite the variance of the total synaptic current $I_\mathrm{syn}=I_\exc+I_\inh$ increasing. Note that the Poisson condition can be relaxed, since it is sufficient for this result that $\frac{\sigma_{\exc,\inh}}{g_{\exc,\inh}^0}\rightarrow0$ \linelabel{lne:poisson}.

\begin{figure*}
\begin{adjustwidth}{-0pt}{0in}
\centering
	\includegraphics[scale=1]{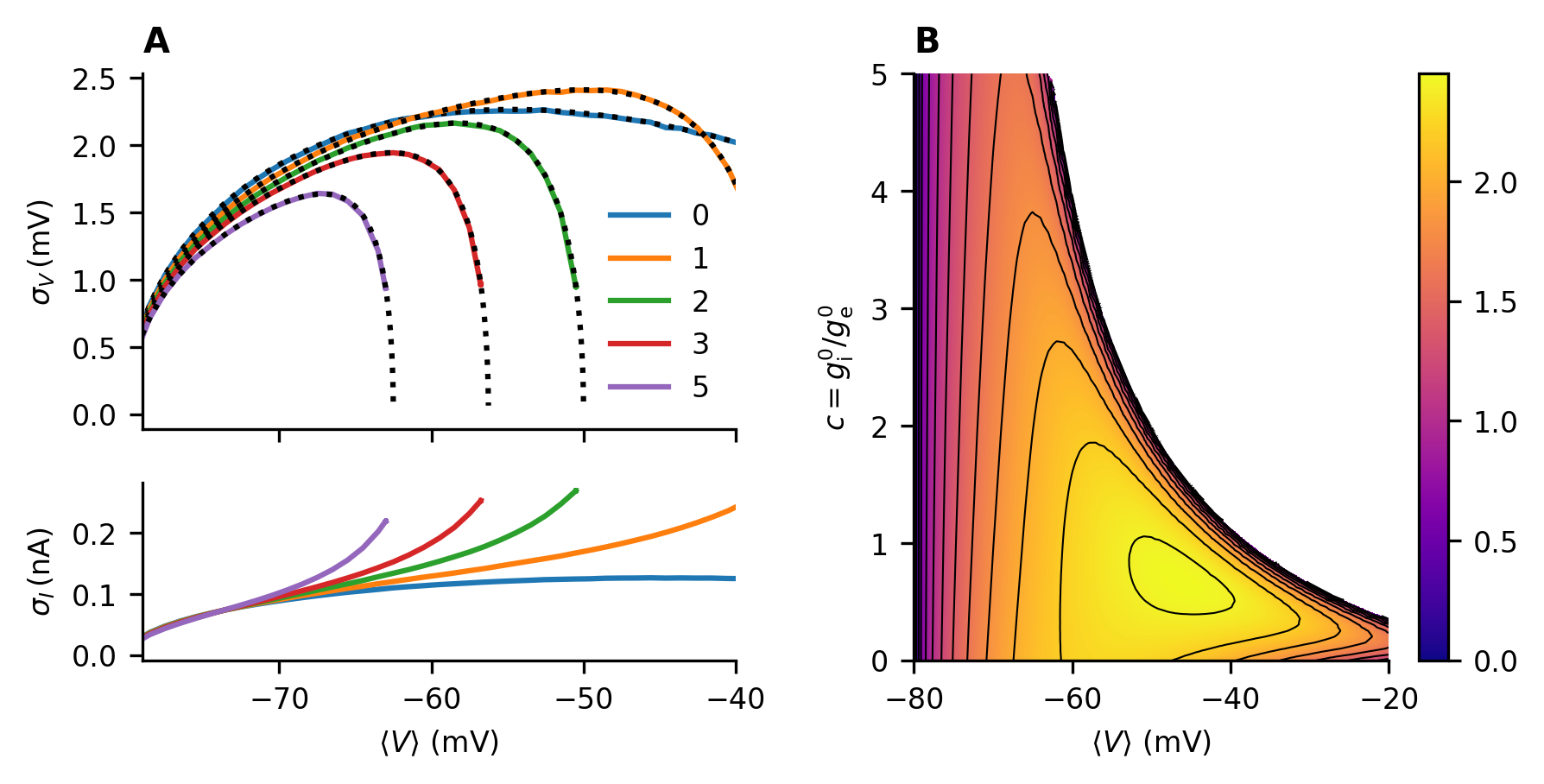}
	\caption{{\bf Stabilization of the membrane potential.} A, top panel: Membrane potential fluctuations as a function of the mean membrane potential for different values of $c$. The full lines represent data obtained from simulations with different excitatory input intensities $\lambda_\exc$. The dotted lines represent the effective time-constant approximation (ETA, Appendix \ref{ap:eta}). Bottom panel: The standard deviation $\sigma_I$ of the total synaptic current $I_\mathrm{syn}=I_\exc+I_\inh$. Note that $\sigma_V$ decreases with growing $c$ even though $\sigma_I$ increases. B: Overview of $\sigma_V$ (color) for all achievable $\langle V \rangle$ ($x$-axis) at given $c$ ($y$-axis). $C=\SI{1}{\micro\farad/\cm^2}$ and $\gleak=\SI{0.045}{\milli\siemens/\cm^2}$, approximation of $\sigma_V$ computed from the ETA. Heatmaps for different values of $\gleak$ are provided in the \suplfigB{} and for different values of $A_\inh$ (Eq. (\ref{eq:expon})) in the \suplfigC{}.}
	\label{fig:mean_std}
\end{adjustwidth}
\end{figure*}

\linelabel{lne:star_explain}Let $\sigma_V(\langle V\rangle; c)$ be the function specifying the standard deviation of the membrane potential with mean $\langle V\rangle$, parametrized by $c$. It is a continuous function, with $\sigma_V(\EL; c)=\sigma_V(V_\infty(c); c)=0$, otherwise $\sigma_V(\langle V\rangle; c)>0$. Note that lower $c$ leads to higher $V_\infty(c)$. Therefore, given $c_1>c_2$, there has to be an interval close to $V_\infty(c_1)$ where $c_2$ results in lower membrane fluctuations. Moreover, simulations indicate that this holds, even in non-limit regimes (Fig \ref{fig:mean_std}A, top panel). \linelabel{lne:highchighinh_start}This result is rather counter-intuitive, since with an increase in $c$, it is necessary to increase both $\lambda_\exc$ and $\lambda_\inh$ (if $\langle V \rangle > \Ei$), and thus simultaneously increase synaptic current fluctuations (Fig \ref{fig:mean_std}A, bottom panel) in order to keep the membrane potential constant\linelabel{lne:highchighinh_end}. With our choice of parameters, lower $c$ may also result in a slight decrease in membrane potential fluctuations. This is mainly due to the membrane time constant $\tau=\frac{C}{\gleak}\doteq\SI{22}{\milli\second}$. The shorter the time constant, the closer $V$ follows $\vef$, and the smaller the region in which decreasing $c$ leads to lower membrane potential variability (see the \suplfigB{})\linelabel{lne:stop_explain}.

\subsection*{Effects on firing regularity}

The regularity of spike-firing is important for information transmission between neurons \cite{Toyoizumi2006,Kostal2007,RuytervanSteveninck1997,Strong1998}. In the previous section, we demonstrated that if appropriate synaptic drive is used, higher inhibitory input rates (or equivalently higher inhibition-to-excitation ratio $c$) lead to lower membrane potential fluctuations. In this section, we focus on the effects of inhibition on post-synaptic firing regularity, particularly on the regularity of a post-synaptic spike train with a fixed frequency evoked by different stimuli with different levels of inhibition.

\subsubsection*{Generalized Leaky Integrate-and-Fire models}

For our analysis, it is essential to distinguish two different input regimes: 1. Sub-threshold regime: $E_0 \leq \theta$ and 2. Supra-threshold regime: $E_0 > \theta$, where $\theta$ is the firing threshold.

In the sub-threshold regime, firing activity is driven by fluctuations in the membrane potential. Therefore, increasing the input rates $\lambda_\exc$, $\lambda_\inh$ and simultaneously keeping $E_0$ constant leads to a decrease in firing rate due to suppressed membrane potential fluctuations (note that an analogous effect was described in the Hodgkin-Huxley model \cite{Tiesinga2000}). In order to maintain the post-synaptic firing rate (PSFR) constant while increasing the input rates, it is necessary to compensate for the decrease in fluctuations by increasing $E_0$. Therefore, it is not intuitively clear whether the decrease in membrane potential fluctuations will lead to an increase in firing regularity.

In the supra-threshold regime, the firing activity is given by the driving force on the membrane potential $(V-\vef)/\tef$. Fluctuations in the interspike intervals are then given mostly by the fluctuations of $\vef$. However, lower fluctuations of $\vef$ are associated with lower $\tef$ and it is necessary to decrease $E_0$, if one wishes to decrease the fluctuations of $\vef$ and keep the firing rate constant at the same time. Intuitively, the fluctuations of $\vef$ will impact the firing regularity more, if the difference $(V-\vef)$ is lower. Therefore it is again unclear how the increased synaptic fluctuations affect the firing regularity.

\begin{figure*}
\begin{adjustwidth}{-0pt}{0pt}
    \centering
	\includegraphics[scale=1]{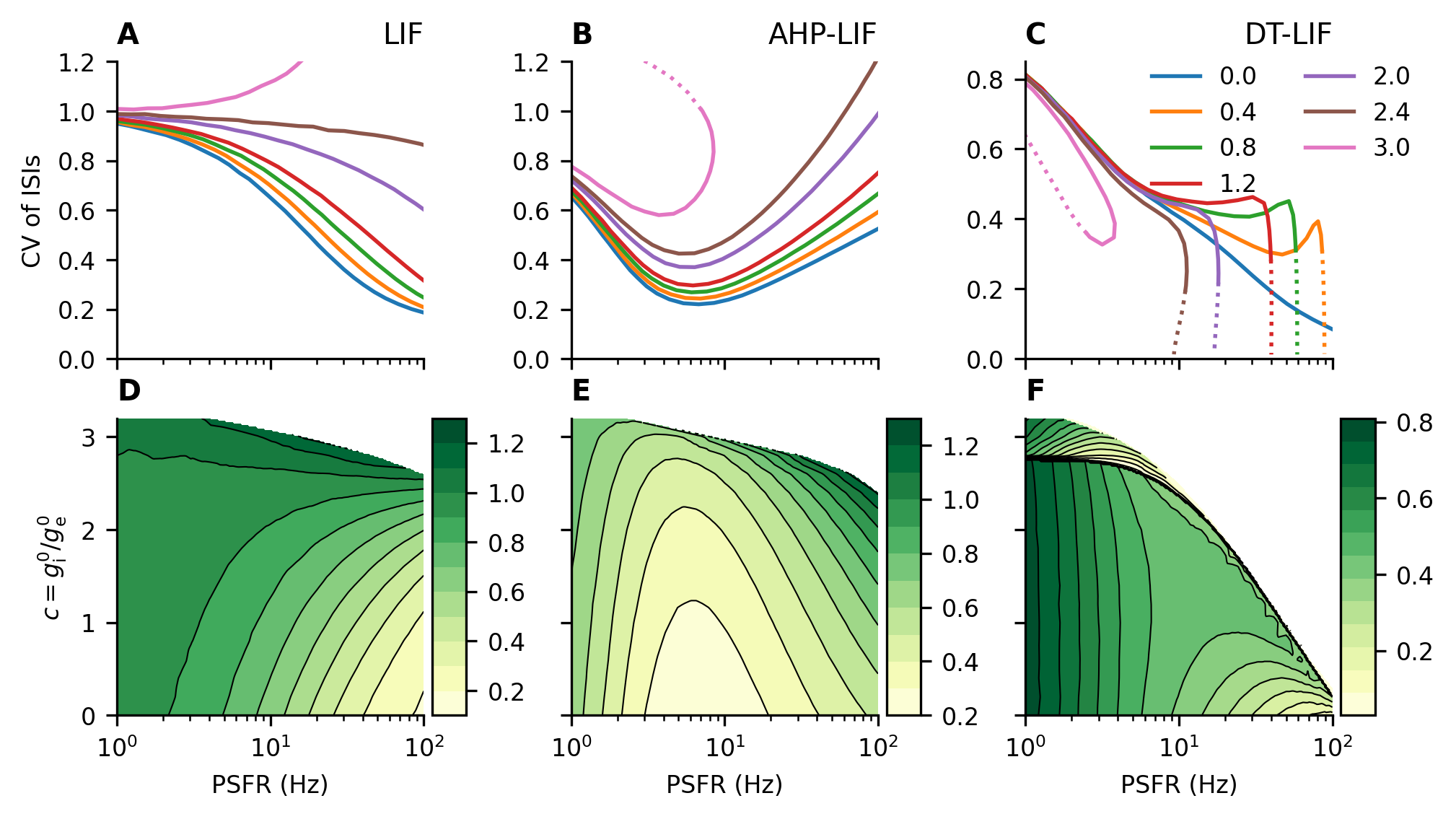}
	\caption{{\bf The effect of membrane potential stabilization on spiking regularity in the GLIF models.} A-C: Dependence of the \cv{} of ISIs on the post-synaptic firing rate for different values of $c$ (color-coded). The dotted parts of the curves represent the sections where $\lambda_\exc>\SI{100}{\kilo\hertz}$. In the LIF and AHP-LIF model, higher $c$ universally leads to higher \cv{}. In contrast, in the DT-LIF model, higher $c$ can lead to more regular spike trains, especially if the input intensities are high. If $V_\infty(c)\leq \theta$ (or $\theta_0$ for the DT-LIF model, i.e., $c\geq2.75$), the firing rate will eventually drop to 0. D-E: Contour plots with color-coded \cv{}, $c$ on the $y$-axis. If more than one input can produce the same PSFR with the same $c$, the lowest possible value of \cv{} is color-coded, resulting in the discontinuity in F. The data points were obtained from simulations with different input intensities $\lambda_\exc$, $\lambda_\inh$.}
	\label{fig:glif}
\end{adjustwidth}
\end{figure*}

In general, we observe that in the suprathreshold regime, the \cv{} of ISIs decreases with growing PSFR (Fig \ref{fig:glif}A,D). Moreover, as we show in the Appendix \ref{app:limit}:
\begin{equation}
    \lim_{\lambda_\exc,\lambda_\inh\rightarrow\infty}\cv{}=0
\end{equation}
However, if the firing rate is held constant, the \cv{} increases with growing $c$. Therefore, an increase in the inhibition-to-excitation ratio decreases firing regularity, despite the stabilizing effect on membrane potential.

With high values of $c$, the \cv{} grows locally with increasing firing rate. This is due to the fact that as $E_0$ is very close to the threshold and the membrane time constant (Eq (\ref{eq:tau})) is very low, the neuron fires very rapidly (bursts) when $\vef$ (Eq (\ref{eq:voltage})) exceeds the threshold but is otherwise silent.

For the AHP-LIF model, no improvements are observed in the firing regularity with increasing $c$ (Fig \ref{fig:glif}B,E). At low firing frequencies, the \cv{} of the AHP-LIF model is generally lower than that in the classical LIF model. This is to be expected given the introduction of negative correlations in subsequent ISIs \cite{Chacron2004,Lindner2005,Farkhooi2011}. However, at higher firing rates, higher $c$ actually leads to a higher \cv{} than that observed in the LIF model. This is due to the fact that in regimes where $\vef$ is always above the threshold in the LIF model, the hyperpolarizing M-current drives the time-dependent effective reversal potential $\vefm$ (Eq (\ref{eq:vefm})) closer to the threshold. This leads to bursting, similar to that observed in the LIF model with $E_0$ near threshold. This is illustrated in more detail in the Appendix \ref{app:limit}, where we also demonstrate that if $V_\infty(c)>V_\mathrm{thr}$, then $\cv{}\rightarrow0$, similar to the LIF model.

In the DT-LIF model, with the limit of infinite conductances, the membrane potential will reach $V_\infty(c)$ immediately after a spike is fired. If $V_\infty(c) \geq \theta_0$, the neuron will fire with exact ISIs
\begin{equation}
	T = \tau_\theta\log\left(1+\frac{\Delta\theta}{V_\infty(c)-\theta_0} \right).
\end{equation}
Therefore, any firing rate lower than $\left(\tau_\theta\log\left(1+\frac{\Delta\theta}{V_\infty(c)-\theta_0} \right)\right)^{-1}$ can be asymptotically reached with $\cv{}=0$. Thus, firing regularity can always be improved by increasing $c$, similar to the case of membrane potential variability. However, very high input intensities are necessary to observe such regularization. Further, with biologically realistic input intensities (excitatory input intensity up to \SI{100}{\kilo\hertz}), increased regularity with higher $c$ is observed only for post-synaptic firing rates below approximately \SI{20}{\hertz} (Fig \ref{fig:glif}C,F).

Note that the structure of the contour plot in Fig \ref{fig:glif}F is very similar to that in Fig \ref{fig:mean_std}B, i.e., approximately for $c>1$, an increase in $c$ stabilizes the membrane potential and increases the spike-firing regularity. The opposite is observed for $c<1$. Moreover, the structure of the heatmap changes accordingly if the membrane time constant is decreased by increasing $\gleak$ (\suplfigB{}) or if the inhibitory synaptic connections are strengthened (\suplfigC{}).

\subsubsection*{Hodgkin-Huxley models}

\begin{figure*}
\begin{adjustwidth}{0pt}{0pt}
\centering
	\includegraphics[scale=1]{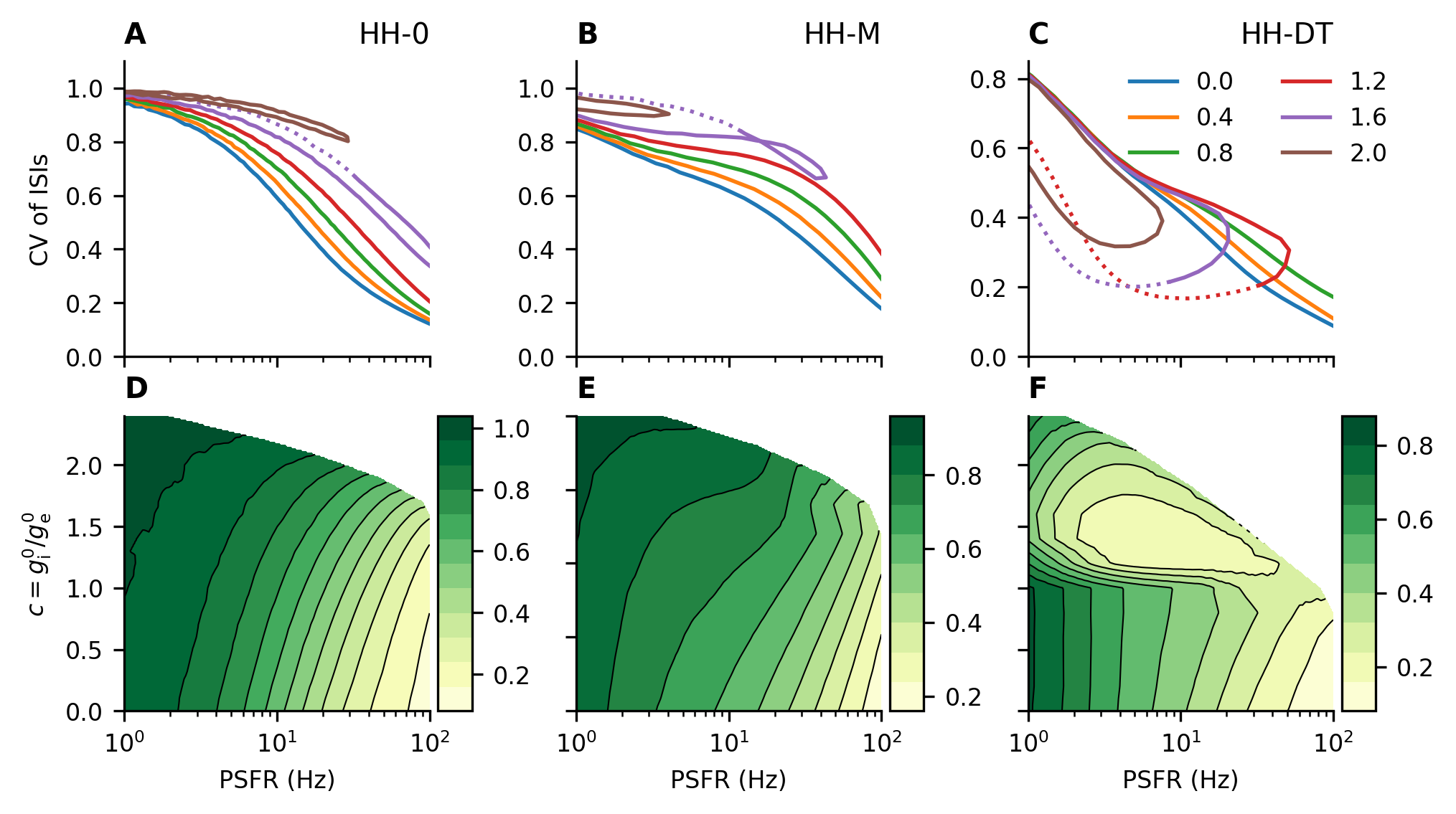}
	\caption{{\bf The effect of membrane potential stabilization on spiking regularity in the Hodgkin-Huxley models.} A-C: Dependence of the \cv{} of ISIs on the firing rate for different values of $c$ (color-coded). The dotted parts of the curves represent the sections where $\lambda_\exc>\SI{100}{\kilo\hertz}$. In the subthreshold regimes, the output rate reaches its maximum and then starts dropping to zero. For the HH-DT model (C), the \cv{} decreases at this point, whereas for the HH-0 (A) and HH-M (B) models, no clear improvement is observed. D-F: Contour plots with color-coded \cv{}, $c$ on the $y$-axis. If more than one input can produce the same PSFR with the same $c$, the lowest possible value of \cv{} is color-coded. Note that there is no discontinuity in F, unlike in Fig. \ref{fig:glif}F. The transition to the more regular states with growing $c$ is continuous, as is illustrated in \suplfigD{}. The data points were obtained from simulations with different input intensities $\lambda_\exc$, $\lambda_\inh$.}
	\label{fig:hh_regularity}
\end{adjustwidth}
\end{figure*}

Generally, the behavior of the HH models is very similar to that of their GLIF counterparts (Fig \ref{fig:hh_regularity}). Similar \textquote{subthreshold} behavior is apparent - for high values of $c$, the firing rate starts dropping to zero with increasing input intensity.

Similarly to the GLIF models, no improvements are observed with growing $c$ for the HH-0 (Fig \ref{fig:hh_regularity}A,D) and HH-M (Fig \ref{fig:hh_regularity}B,E) models. For the HH-DT model, lower \cv{} of ISIs can always be achieved in the subthreshold regime, when the rate starts dropping back to zero due to the strong input (Fig \ref{fig:hh_regularity}C,F).

\begin{figure}
\begin{adjustwidth}{0pt}{0pt}
\centering
	\includegraphics[scale=1]{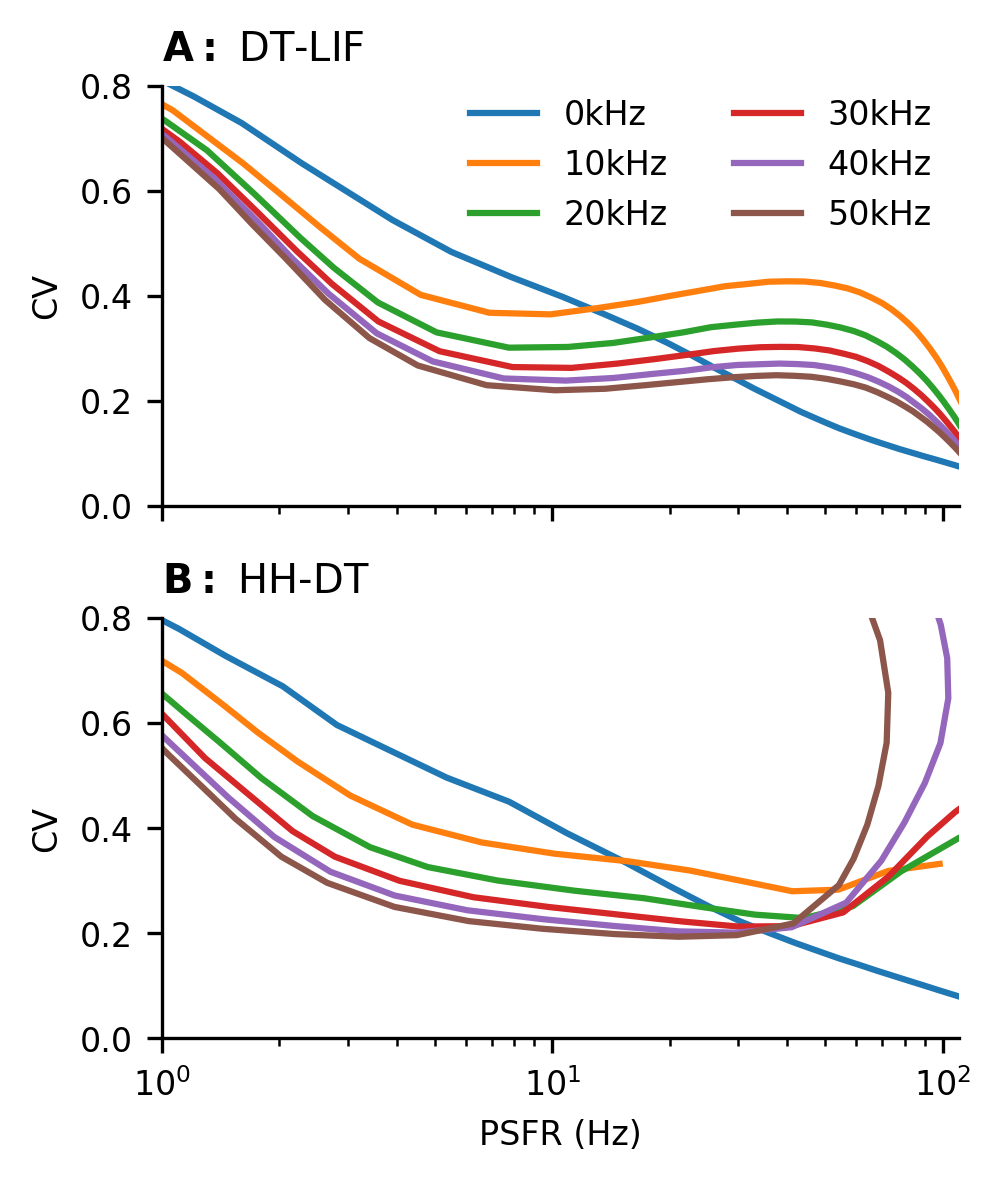}
	\caption{{\bf Constant inhibition trajectories for the dynamic threshold models.} In both the DT-LIF (A) and HH-DT (B) models, increasing the pre-synaptic inhibitory firing rate (color) is beneficial for the firing regularity (measured by \cv{}, $y$-axis) for a wide range of PSFRs ($x$-axis).}
	\label{fig:hh_const_inhrate}
\end{adjustwidth}
\end{figure}

Increasing $c$ in the HH-DT subthreshold regime decreases the \cv{}. However, it is important to note that increased $c$ does not imply stronger inhibitory input in this case. In fact, increasing the inhibitory input rate $\lambda_\inh$ is almost always beneficial for the spike-firing regularity in the HH-DT model, and this is also the case in the DT-LIF model (Fig \ref{fig:hh_const_inhrate}). From this, we conclude that if a neuron exhibits a dynamic threshold, a stimulus will produce a more regular spike train if it elicits an increase in inhibitory input simultaneously with excitatory input.

% Results and Discussion can be combined.
\section*{Discussion}
\subsection*{Simplified input models}
\subsubsection*{Absence of reversal potentials}
If the reversal potentials are not taken into account, the synaptic currents are given by
\begin{equation}
    I_{\exc,\inh}(t) = \sum_{t_k\in\mathcal{T}_{\exc,\inh}}H_{\exc,\inh}(t-t_k),
\end{equation}
where $H$ is again a filtering function. If the two currents are uncorrelated, they will add up to an input current with mean value $I_0$ and standard deviation $\sigma_I$. If the diffusion approximation is employed (the current is modeled as an Ornstein-Uhlenbeck process with a time constant $\tau_I$), the mean and standard deviation of the membrane potential are \cite{Lindner2006b}:
\begin{align}
    \langle V\rangle_I &= \EL+\frac{I_0}{\gleak}, \\
    \sigma_{V,I}^2 &= \sigma_I^2\frac{\tau_I}{a^2\gleak(C+\gleak\tau_I)}.
\end{align}
In the absence of synaptic reversal potentials, the variance diverges with growing input, and increasing the synaptic current fluctuations by increasing $\lambda_\exc$ and $\lambda_\inh$ clearly increases the membrane potential fluctuations, in contrast to the model with synaptic reversal potential.

\subsubsection*{Absence of synaptic filtering}
\begin{figure*}
\begin{adjustwidth}{0pt}{0pt}
\centering
	\includegraphics[scale=1]{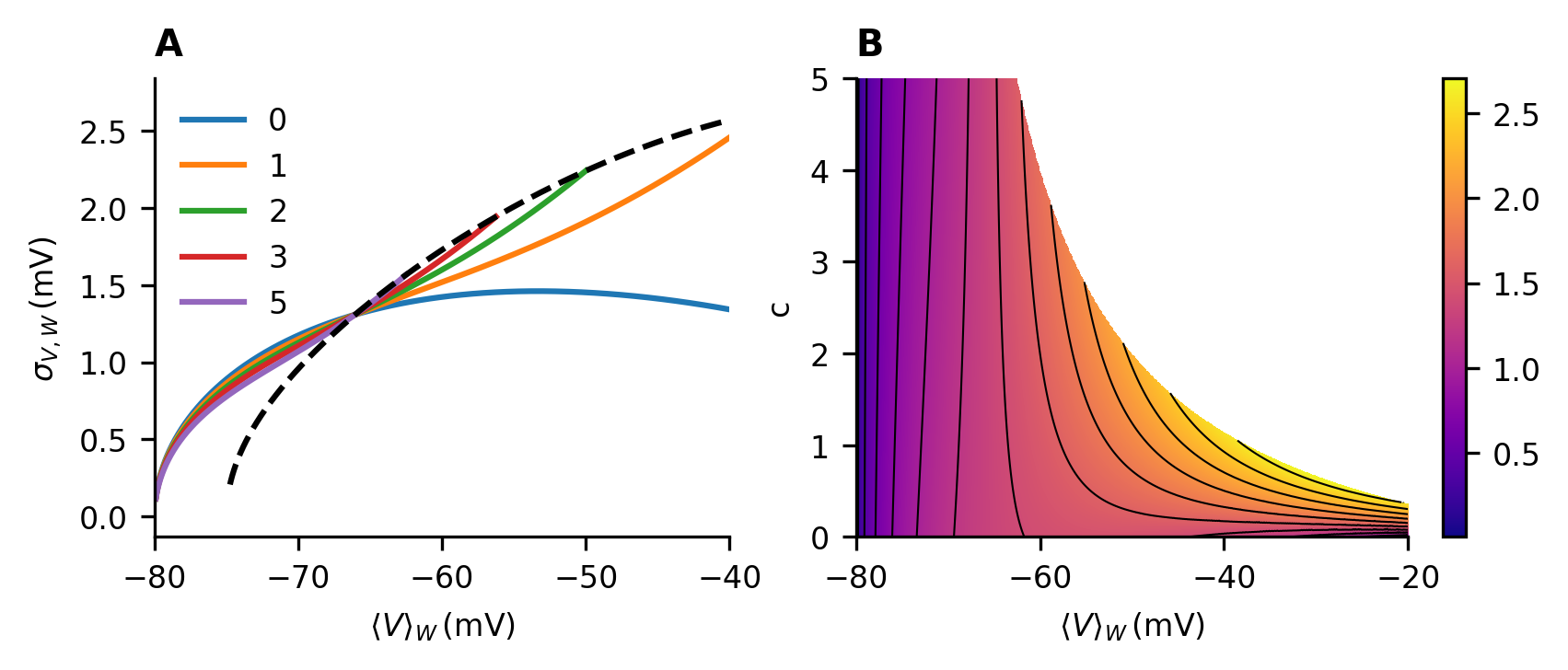}
	\caption{{\bf Membrane potential with conductance input without synaptic filtering.} A: Membrane potential fluctuations as a function of the mean membrane potential for different values of $c=\frac{\lambda_\inh b_\inh}{\lambda_\exc b_\exc}$ (color-coded), as calculated from the Eqs (\ref{eq:E_rich},\ref{eq:std_rich}). The dashed line represents the limit $\lim\limits_{\lambda_\exc,\lambda_\inh\rightarrow\infty}\sigma_{V,W}(\langle V\rangle_{W})$. The membrane potential is not stabilized at infinite inputs. Above certain depolarizations, inhibition increases membrane potential fluctuations, contrary to the case of conductance input with synaptic filtering. B: Heatmap with color-coded standard deviation of the membrane potential. Parameters used were $b_\exc=0.0045$, $b_\inh=0.0150$.}
	\label{fig:whitenoise}
\end{adjustwidth}
\end{figure*}

If synaptic filtering is neglected, $h_{\exc,\inh}$ become $\delta$-functions:
\begin{equation}
	h_{\exc,\inh}(t) = Ca_{\exc,\inh}\delta(t),
	\label{eq:deltafunc}
\end{equation}
where $C$ is the membrane capacitance, and $a_{\exc,\inh}$ governs the jump in the membrane potential $\Delta V$ triggered by a single pulse:
\begin{equation}
	\Delta V = (E_{\exc,\inh}-V)(1-e^{-a_{\exc,\inh}}).
\end{equation}
This model was studied extensively, e.g., in \cite{Lanska1994,Richardson2004,Richardson2006}. In \cite{Richardson2004, Richardson2006}, the formulas for the mean membrane potential and its standard deviation are calculated in the diffusion approximation:
\begin{align}
	\langle V\rangle_{W} &= \tau(\EL\tau_L^{-1}+\Ee\lambda_\exc b_\exc + \Ei\lambda_\inh b_\inh)\label{eq:E_rich} \\
	\sigma_{V,W}^2 &= \frac{\tau_L}{2}\frac{\lambda_\exc b_\exc^2(\langle V\rangle-\Ee)^2+\lambda_\inh b_\inh^2(\langle V\rangle-\Ei)^2}{1+\tau_L\lambda_\exc b_\exc(1-b_\exc/2)+\tau_L\lambda_\inh b_\inh(1-b_\inh/2)}, \label{eq:std_rich}
\end{align}
where
\begin{align}
	\tau^{-1} &= \tau_L^{-1}+\lambda_\exc b_\exc+\lambda_\inh b_\inh \\
	b_{\exc,\inh} &= 1-e^{-a_{\exc,\inh}}.
\end{align}
Richardson \cite{Richardson2004} reported that a higher inhibition-to-excitation ratio may lead to a decrease in the membrane potential fluctuations for strongly hyperpolarized membranes. However, the effect of inhibition reverses as the membrane potential depolarizes (Fig \ref{fig:whitenoise}). Furthermore, the membrane potential does not stabilize within the limit of infinite firing rates. Therefore, the time correlation of synaptic input introduced by synaptic filtering is necessary to observe the shunting effect of inhibitory synapses.

\subsection*{Regular firing in multicompartmental models}

The models analyzed in this work are all single-compartmental models, i.e., models in which the charge is distributed infinitely fast across the cell, and the membrane potential is therefore the same everywhere. In reality, neurons receive input predominantly at dendrites, and the spikes are initiated in the soma. To account for this fact, multicompartmental models are typically employed. The soma and dendritic parts can be modeled as two separate compartments (for simplicity, as two identical cylinders) connected through a coupling conductance $g_c$ :
\begin{align}
	C\oder{V_S}{t} &= -\gleak(V_S-\EL)-g_c(V_S-V_D) \\
	\begin{split}
	C\oder{V_D}{t} &= -\gleak(V_D-\EL)-g_c(V_D-V_S)-\\&-\frac{1}{a_D}(\gexc(V_D-E_\exc)+\ginh(V_D-E_\inh))
	\end{split}
\end{align}
where $V_S$ and $V_D$ are the membrane potentials of the somatic and dendritic compartments, respectively; $a_D$ is the dendritic area; and $V_S$ is reset to $V_r$ when the threshold $\theta$ is reached.

In the hypothetical case of infinite input rates, $V_D=V_\infty(c)$ and $V_S$ periodically decay to $V_S^0=\frac{\gleak \EL+g_c V_D}{\gleak+g_c}$ with a time constant $\tau_2=\frac{\gleak+g_c}{a_S C}$, resulting in regular ISIs
\begin{equation}
	T = -\tau_2\log\left( 1+\frac{V_r-\theta}{V_S^0-V_r} \right).
\end{equation}
Therefore, it is possible to reach a wide range of firing rates with $\cv{}=0$ and decrease \cv{} while maintaining a constant mean firing rate by increasing $c$, similar to the case of LIF with a dynamic threshold.

The coupling conductance can be calculated as $g_c=\frac{d}{4R_a l^2}$ \cite{SterrattPrinciples2011}, where $d$ is the diameter of the cylinder, $l$ is the length, and $R_a=\SI{150}{\ohm\cm}$ is the longitudinal resistance. If we consider that the original area of the neuron approximately $\SI{3.5e-4}{\cm^2}$ is split between the two cylinders and we set $d=l$, we obtain $\tau_2\approx\SI{4.5}{\micro\second}$. It is therefore unlikely that firing rate regularization with biologically relevant post-synaptic firing rates would be observed with biologically plausible inputs.

\section*{Conclusion}

We demonstrate that a higher inhibition-to-excitation ratio and subsequently higher synaptic current fluctuations lead to a more stable membrane potential if the stimulation is modeled as time-filtered activation of synaptic conductances with reversal potentials. Our analysis thus provides a theoretical context for the experimental observations of \cite{Monier2003}. Moreover, our results highlight the importance of incorporating synaptic filtering and reversal potentials into neuronal simulations. The qualitative differences between neurons stimulated with white noise and colored noise current have been reported in the literature \cite{Brunel2001,Fourcaud2002,Moreno-Bote2004}. However, we demonstrate that realistic synaptic filtering with reversal potentials is responsible for a novel fluctuation-stabilization mechanism which cannot be observed in simplified models.

We analyzed the impact of membrane potential stabilization on spike-firing regularity in GLIF models and HH-type models. We compared the effects of an increased inhibition-to-excitation ratio on two different mechanisms of spike-firing adaptation: adaptation by a hyperpolarizing ionic current (AHP and M-current adaptation) and adaptation implemented as a dynamic firing threshold. Both SFA mechanisms are biologically relevant and are useful in neuronal modeling \cite{Teeter2018,Benda2010,Chacron2000,Kobayashi2009,Kobayashi2016,Levakova2019,Gerstner2009,Kobayashi2019}. We demonstrated that while an increase in inhibition leads to less regular spike trains in the ionic current adaptation models and models without any spike-firing adaptation, it may enhance the firing regularity in the dynamic threshold models. We observed this effect in both the GLIF models and HH-type models.

High presynaptic inhibitory activity is typical of cortical neurons. In the so called high-conductance state, total inhibitory conductance can be several-fold larger than total excitatory conductance \cite{Destexhe2003}. Our findings therefore provide a novel view of the importance of the high-conductance state and inhibitory synapses in biological neural networks.

\section*{Acknowledgements}
This work was supported by Charles University, project GA UK No. 1042120 and the Czech Science Foundation project 20-10251S.

\appendix
\section{Effective time constant approximation}
\label{ap:eta}
The Eqs (\ref{eq:mat_pot},\ref{eq:isyn}) can be rewritten as:
\begin{equation}
    Ca\oder{V}{t}=-g_0(V-E_0)-\gexc^F(V-\Ee)-\ginh^F(V-\Ei),
\end{equation}
where $g_0=a\gleak+\gexc^0+\ginh^0$, $g_{\exc,\inh}^0$, and $g_{\exc,\inh}^F$ are the mean and fluctuating parts of the conductance input. The input can then be separated into its additive and multiplicative parts:
\begin{equation}
    \gexc^F(V-\Ee)=\gexc^F(E_0-\Ee) + \gexc^F(V-E_0).
\end{equation}

By neglecting the multiplicative part $\gexc^F(V-E_0)$, we obtain the effective time constant approximation (ETA). In the diffusion approximation, the mean and standard deviation of the membrane potential are \cite{Richardson2004,Richardson2005}:
\begin{align}
    \langle V \rangle_{\mathrm{ETA}} &= E_0, \\
    \begin{split}
    \sigma_{V,\mathrm{ETA}}^2 &= \left( \frac{\sigma_\exc}{g_0}
    \right)^2(\Ee-E_0)\frac{\tau_\exc}{\tau_\exc+\tau_0}+\\
    &+\left( \frac{\sigma_\inh}{g_0} \right)^2(\Ei-E_0)\frac{\tau_\inh}{\tau_\inh+\tau_0},
    \end{split}
\end{align}
where $\tau_0=\frac{aC}{g_0}$ is the effective time constant, and $\sigma_{\exc,\inh}$ are the standard deviations of the excitatory and inhibitory inputs.

\section{Limit cases of LIF and AHP-LIF models}
\label{app:limit}
\subsection*{High conductance limit of the LIF model}

\begin{figure}
\centering
	\includegraphics[scale=1]{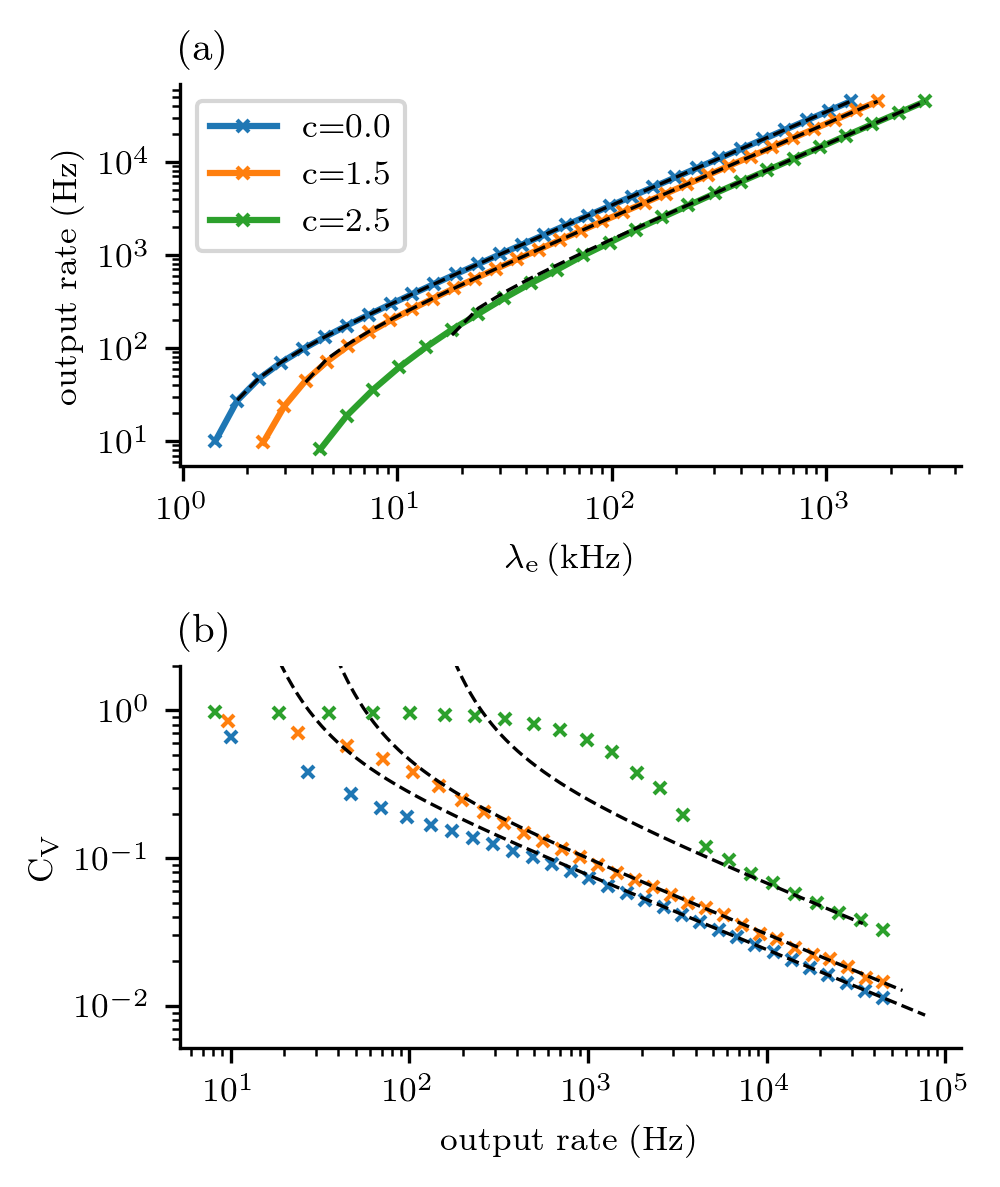}
	\caption{{\bf Approximation of the PSFR and \cv{} of LIF.} Simulation results are color-coded. The dashed lines represent the approximations from the Eqs (\ref{eq:lif_approx_mean},\ref{eq:lif_approx_std}). The firing rate is approximated very well (A). \cv{} is approximated well for very high PSFRs ($>\SI{1}{\kilo\hertz}$) (B). This is due to the short input time constants (\SI{3}{\milli\second} for excitatory, \SI{10}{\milli\second} for inhibitory). For the simulation, we used the timestep $\Delta t=\SI{0.1}{\milli\second}$ if the expected PSFR was $<\SI{100}{\hertz}$ and $\Delta t \frac{\avgisi}{\SI{10}{\milli\second}}$ otherwise.}
	\label{fig:lif_var}
\end{figure}

If $V_\infty(c)>\theta$, in the case of high input intensities, $\vef$ is permanently above the threshold, and the effective membrane time constant $\tau(\gexc,\ginh)$ approaches zero. Therefore, in the absence of a refractory period, the firing rate $f=\frac{1}{\avgisi}$ diverges ($\avgisi$ is the average ISI). If the average postsynaptic ISI is much shorter than synaptic timescales, we can assume that the input remains effectively constant during the entire ISI (corresponding to the adiabatic approximation \citep{Moreno-Bote2004,Moreno-Bote2005,Moreno-Bote2006,Moreno-Bote2008,Moreno-Bote2010}). The length of the ISI is then determined solely by the immediate values of the excitatory and inhibitory conductances
\begin{equation}
    T(\gexc,\ginh) = -\frac{aC}{a\gleak+\gexc+\ginh} \log\left(\frac{\theta-\vef(\gexc,\ginh)}{V_r-\vef(\gexc,\ginh)}\right).
\end{equation}
Assuming independence of the inputs, the mean ISI and its standard deviation can then be approximated as
\begin{gather}
    	\avgisi =  -\frac{aC}{\gtot} \log\left(\frac{\theta-E_0}{V_r-E_0}\right), \label{eq:lif_approx_mean}\\
        \stdisi^2 = \left(\frac{\partial T}{\partial \gexc}\right)^2\Bigr|_{\substack{\gexc=\gexc^0}}\sigma_\exc^2 + \left(\frac{\partial T}{\partial \ginh}\right)^2\Bigr|_{\substack{\ginh=\ginh^0}}\sigma_\inh^2 = \label{eq:lif_approx_std} \\
	= (aC)^2 \gexc^0 \left[ \frac{A_\exc \left( (E_\exc-E_0)(\theta-V_r)+\alpha \right)}{\gtot^4(\theta-E_0)^2(E_0-V_r)^2}\right. +\nonumber\\
	+ \left.\frac{c A_\inh \left( (E_\inh-E_0)(\theta-V_r)+\alpha \right)}{\gtot^4(\theta-E_0)^2(E_0-V_r)^2}\right],\nonumber\\
	\alpha = (\theta-E_0)(E_0-V_r)\log\frac{E_0-\theta}{E_0-V_r},\nonumber
\end{gather}
where $\gtot=a\gleak+\gexc^0+\ginh^0$ (for validity of the approximation see Fig. \ref{fig:lif_var}).
Therefore, $\stdisi/\avgisi=O\left(\left(\gexc^0\right)^{-1/2}\right)$. We conclude that with growing input intensity, the firing rate diverges and $\cv{}\rightarrow0$.

\subsection*{High conductance limit of the AHP-LIF model}
\label{s:musc}

\subsubsection*{Effective reversal potential}
\label{ss:musc_rev}

We follow the assumption that the fluctuations in $\vef(\gexc,\ginh)$ are very small and therefore $\vefm$ (Eq. \eqref{eq:vefm}) is permanently above the threshold $\theta$. With the ISI $\avgisim\ll\tau_\AHP$, $\gahp(t)\approx\langle \gahp(t) \rangle = \frac{\Delta g\tau_\AHP}{\avgisi^\AHP}$. Analogously to the Eq. \eqref{eq:lif_approx_mean}), we can use the following implicit equation to approximate the mean ISI:

\begin{equation}
    \avgisim=-\frac{aC}{\gtotm}\log\frac{\theta-E_0^\AHP}{V_r-E_0^\AHP}, \label{eq:musc_interspike},
\end{equation}
where
\begin{align}
    \gtotm &= a\gleak+\gexc^0+\ginh^0+\frac{\Delta g\tau_\AHP}{\avgisi^\AHP}, \\
    E_0^\AHP &= \frac{a\gleak \EL+\gexc^0\Ee+\ginh^0\Ei+\frac{\Delta g\tau_\AHP}{\avgisi^\AHP} \EK}{\gtotm}.
\end{align}

We continue to evaluate the high-conductance limit of $E_0^\AHP$:
\begin{equation}
\begin{gathered}
    \limg E_0^\AHP=V_\infty^\AHP(c) \equiv \\
    \equiv \limg\frac{\frac{a\gleak}{\gexc^0} \EL+ \Ee+c \Ei+\frac{\tau_\AHP}{\gexc^0\avgisim}\Delta g \EK}{ \frac{a\gleak}{\gexc^0} + 1+c+\frac{\tau_\AHP}{\gexc^0\avgisim}\Delta g }.
\end{gathered}
\end{equation}
Clearly, $\frac{a\gleak}{\gexc^0}\rightarrow 0$. Therefore, it is important to evaluate the limit $A=\limg\gexc^0\avgisim$. Then:
\begin{equation}
    V_\infty^\AHP(c)=\frac{\Ee+c \Ei+\frac{\tau_\AHP}{A}\Delta g \EK}{ 1+c+\frac{\tau_\AHP}{A}\Delta g }.
\label{eq:muscelim}
\end{equation}
By multiplying both sides of Eq. (\eqref{eq:musc_interspike}) with $\gtotm$ and then taking the limit of both sides of the equation, we obtain:
\begin{gather}
    a\gleak\avgisi+\gexc^0(1+c)\avgisi+\tau_\AHP\Delta g = -aC\log\frac{\theta-V_\infty^\AHP}{V_r-V_\infty^\AHP}, \\
    A(1+c) + \tau_\AHP\Delta g = -aC\log\frac{\theta-V_\infty^\AHP}{V_r-V_\infty^\AHP}. \label{eq:A}
\end{gather}

\begin{figure*}
\begin{adjustwidth}{0pt}{-27.5pt}
\centering
	\includegraphics[scale=1]{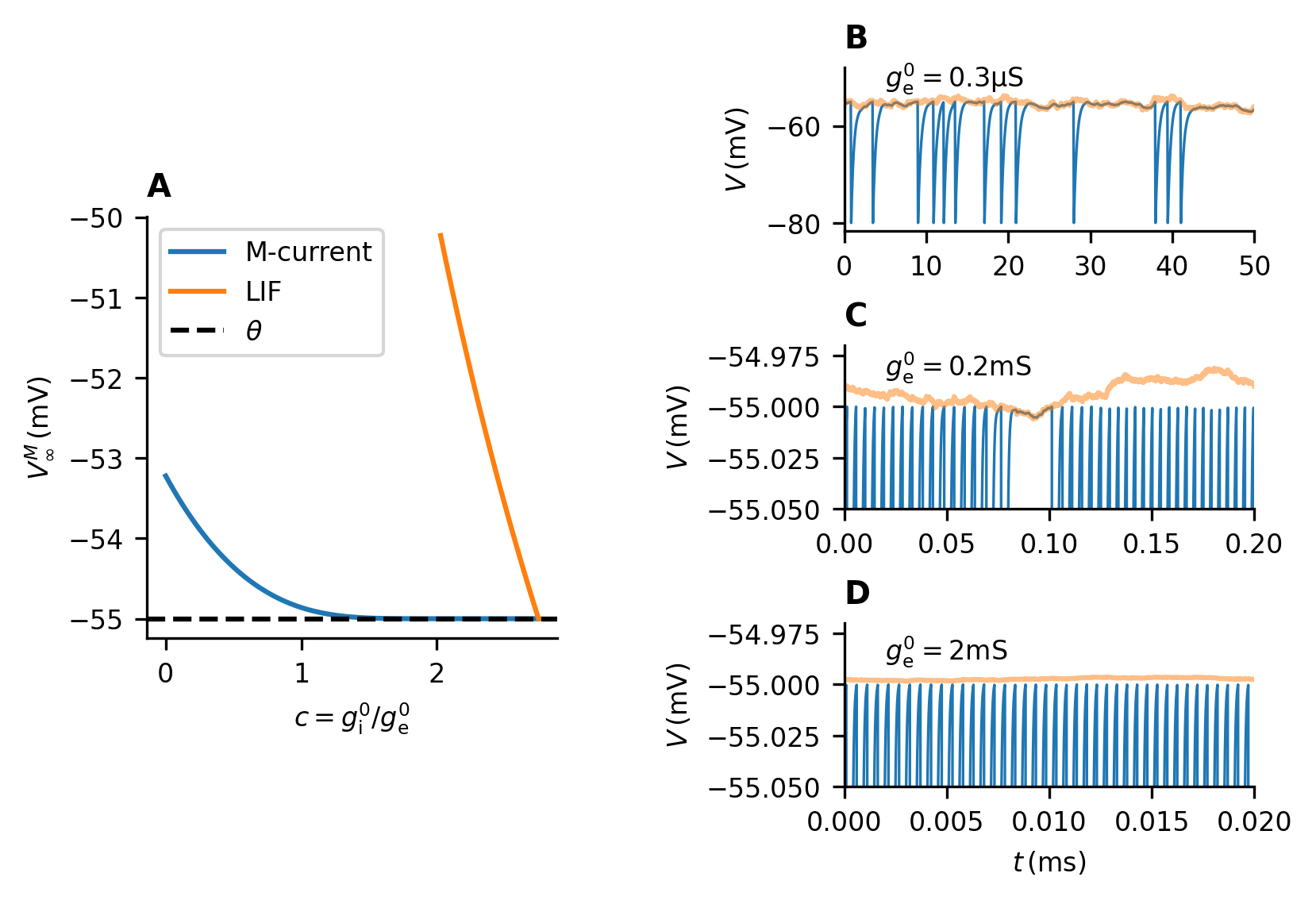}
	\caption{{\bf High conductance limit of the AHP-LIF model} A: Equilibrium potential (in mV) in the infinite conductance limit (Eq. \eqref{eq:muscelim}) for different values of $c$ (shown in blue). For high $c$, the value $V_\infty^\AHP(c)$ is very close to the threshold (dashed black line). The value of $V_\infty(c)$ \eqref{eq:vinfty}, corresponding to the LIF model, is shown in orange for comparison. The M-current adaptation clearly pushes the equilibrium potential closer to the threshold, leading to bursting behavior. B-D: Time-course of the membrane potential of LIF with M-current adaptation with $c=1.7$ for different values of input intensities. The membrane potential (shown in blue) follows closely $\vefm$ (shown in orange). When $\vefm>\theta$, the neuron is bursting; otherwise, the neuron is silent. With higher input intensities, the probability of $\vefm\leq\theta$ drops, and the firing rate becomes increasingly more regular.}
	\label{fig:musc_limit}
\end{adjustwidth}
\end{figure*}

Numerical solution of Eq. (\eqref{eq:A}) allows us to compare $V_\infty^\AHP(c)$ with $V_\infty(c)$ and thus provides a comparison between the LIF model with and without the M-current adaptation (Fig. \ref{fig:musc_limit}). For approximately $c>5$ (with the used parameters), $V_\infty(c)^\AHP\approx\theta$. Therefore, the neuron requires a very high input intensity for the fluctuations to be so small that $\vefm$ permanently exceeds the threshold, and in the range of biologically feasible inputs, the fluctuations in $\vefm$ lead to bursting when $\vefm>\theta$ and are silent when $\vefm\leq\theta$ (Fig. \ref{fig:musc_limit}).

\subsubsection*{The limit of \cv{}{}}
\label{ss:musc_lim}

Neglecting the variance of $\gahp(t)$, the variance of ISIs can then be approximated analogously to Eq. (\eqref{eq:lif_approx_std}) as:
\begin{equation}
    \sigma_\mathrm{ISI}^2=
    \left(\frac{\partial\avgisi}{\partial \gexc}\right)^2\sigma_\exc^2 + \left(\frac{\partial\avgisi}{\partial \ginh}\right)^2\sigma_\inh^2,
\end{equation}
Our goal is to demonstrate that the coefficient of variation (\cv{}) approaches zero. Using the definition of \cv{} (Eq. \eqref{eq:cv}) and Eq. \eqref{eq:musc_interspike}, we have
\begin{gather}
    \limg\cv{} = \limg \frac{\stdisi}{\avgisi} \\
    = -\limg\frac{\gtotm}{aC}\frac{\stdisi}{\log\frac{\theta-E_0^\AHP}{V_r-E_0^\AHP}} \\
    =-\limg \gtotm\stdisi \limg \left( \log\frac{\theta-E_0^\AHP}{V_r-E_0^\AHP} \right)^{-1}.
\end{gather}
Since $-\infty<\limg \left( \log\frac{\theta-E_0^\AHP}{V_r-E_0^\AHP} \right)^{-1}<0$, it remains to be shown that $\limg \gtotm\stdisi=0$, which can be shown by using the implicit differentiation formula.

% \bibliography{references.bib}


\ifarXiv
    

\section*{Supplementary material}

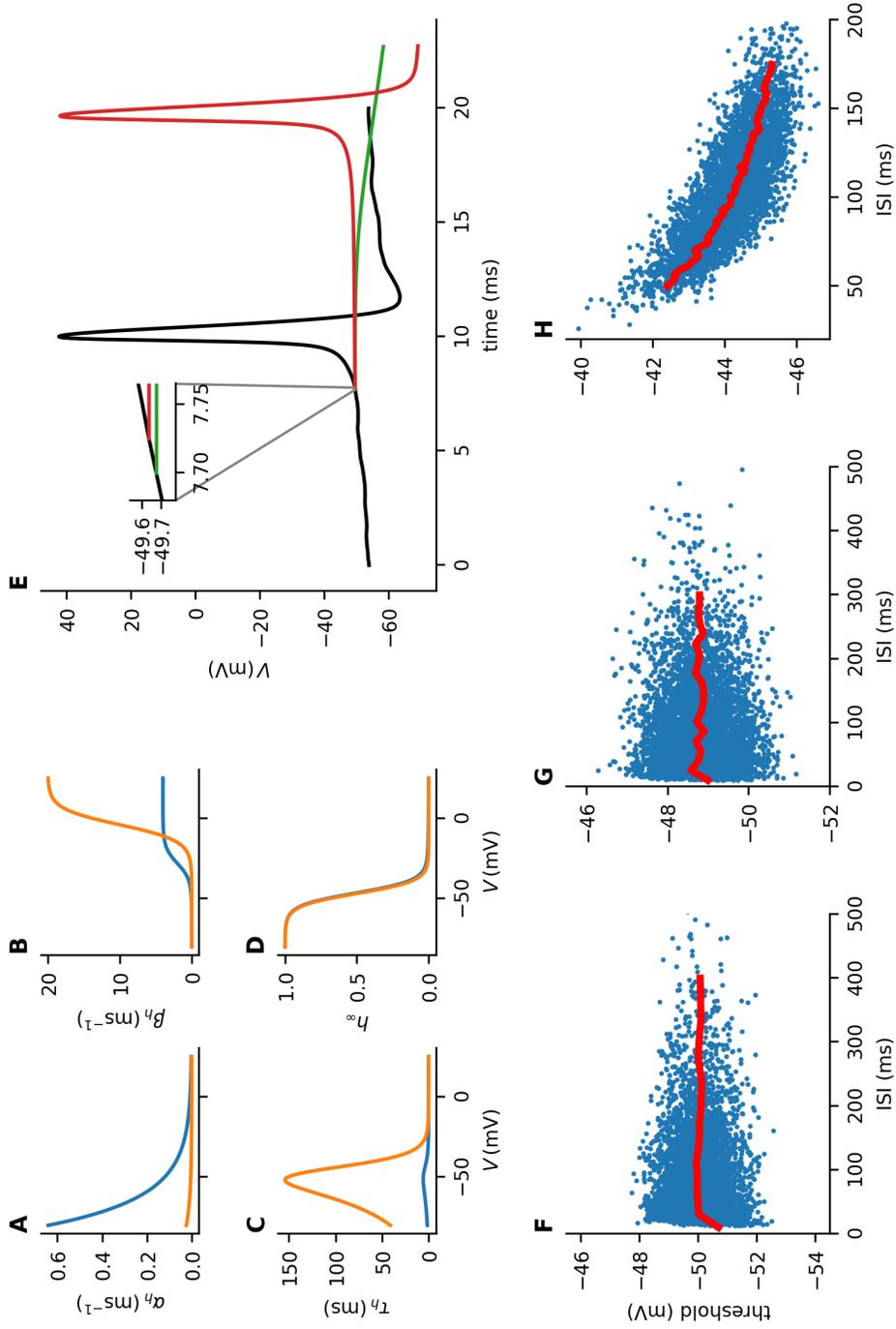

**Supplementary Figure 1: Construction and validation of dynamic threshold HH model.** A-D: The blue lines represent the activation (A) and inactivation (B) rate functions of the gating variable $h$ of the non-adapting HH model (HH-0). The orange lines represent the modified rate functions for the HH-DT model. As an effect of this modification, the recovery time in the HH-DT model is prolonged (C), while the steady state activation $h_\infty$ remains relatively unchanged (D). E: Computation of the threshold. If no additional pre-synaptic spikes arrive after $t = 7.70\,\text{ms}$, no spike is observed (green trace). If the pre-synaptic spikes stop arriving $\Delta t = 0.025\,\text{ms}$ later, a spike is observed. Therefore, we define the threshold as being reached at $t = 7.725\,\text{ms}$. F-H: Comparison of the calculated threshold with the preceding interspike interval (ISI). We validated that the threshold decreases with time after a spike is fired by estimating the firing threshold individually for each spike during a simulation with a stationary input, resulting in a post-synaptic firing rate of approximately 10 Hz Neither the non-adapting model (HH-0, panel F) nor the model with M-current adaptation (HH-M, panel G) displays signs of significant dependence between the threshold and ISI. For the dynamic threshold model (HH-DT, panel H) we observed a clear decrease in the threshold with increasing duration of the preceding ISI. The red line, acting as a visual aid, was obtained by a KNN regression.

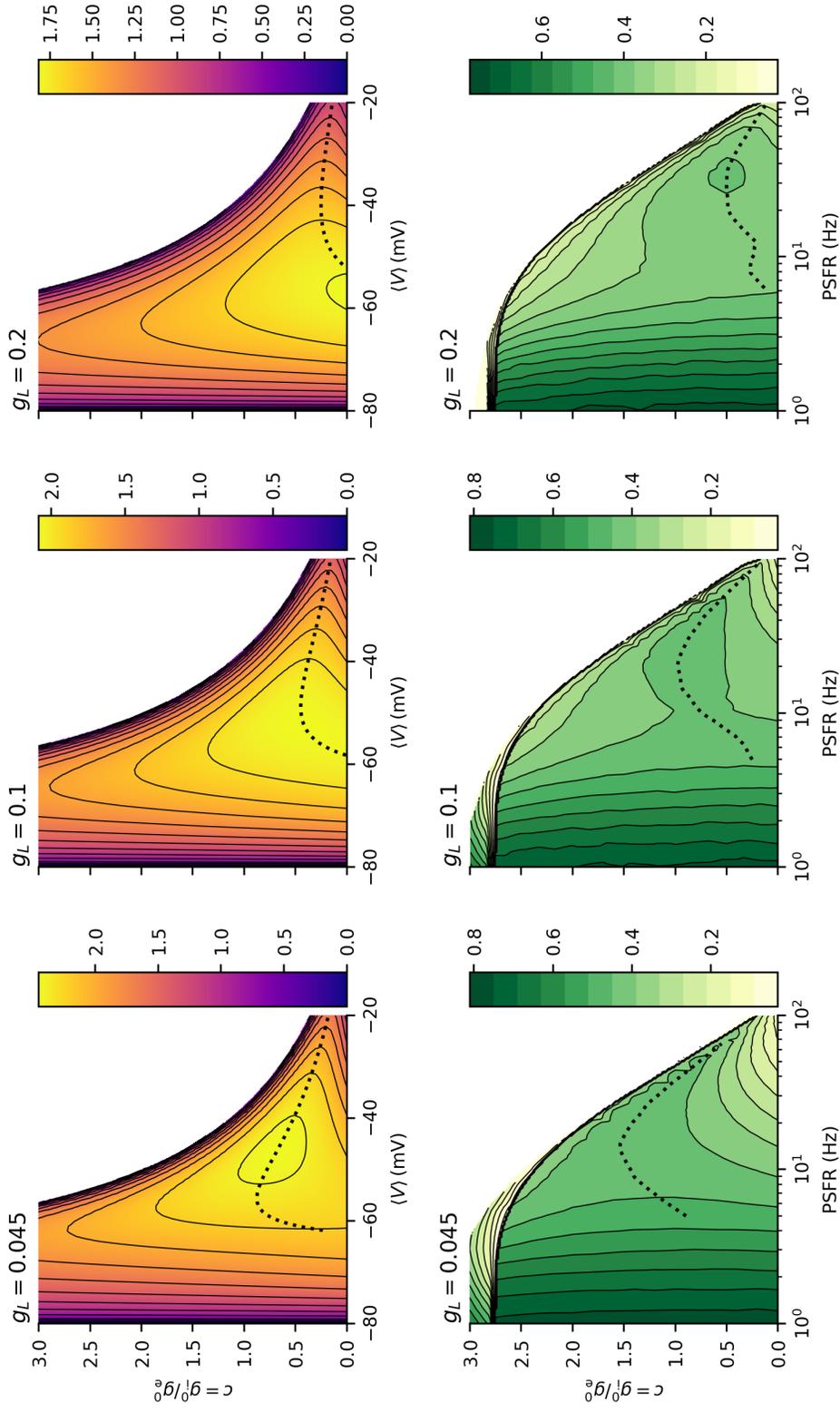

**Supplementary Figure 2: Varying the membrane time constant.** Top row: The membrane time constant is varied by changing the leak conductance $g_L$. With greater leak conductance, the membrane potential $V$ follows $V_{ef}$ more closely and then higher $c$ leads to lower membrane potential fluctuations more universally (approximation of $\sigma_V$, calculated from ETA, is colorcoded). The regions where higher $c$ decreases / increases the membrane potential are separated by the dotted line. Bottom row: The structure of the contour plots indicating how firing regularity ($C_V$, colorcoded) changes with $c$ and PSFR follows the structure of the membrane potential fluctuation heatmaps (top row). With higher $g_L$, higher $c$ increases the firing regularity for a wider range of firing rates and a wider range of $c$. The dotted line approximates the separation between the regions where higher $c$ leads to higher / lower firing regularity.

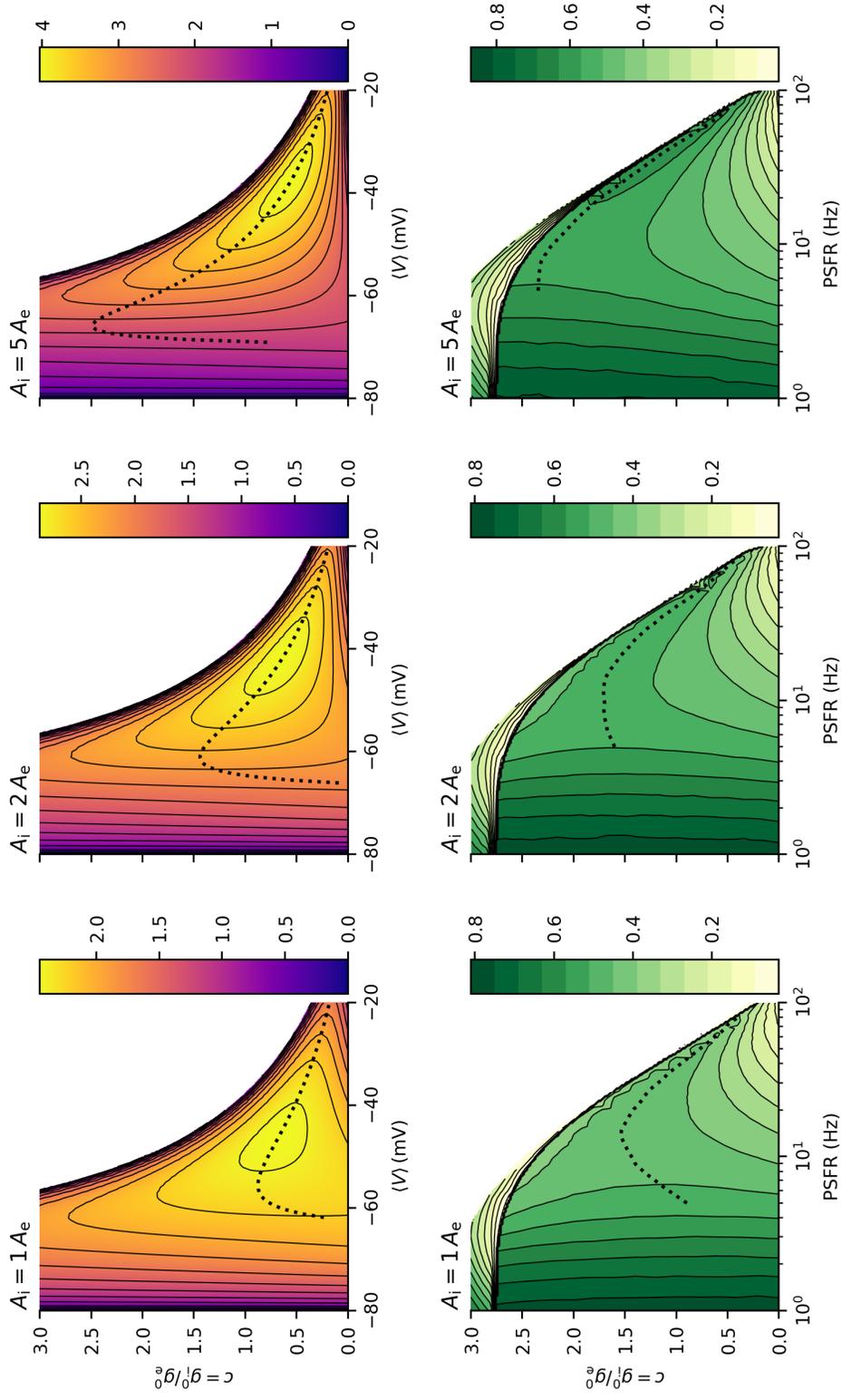

**Supplementary Figure 3: Varying the aplitude of inhibitory synapses.** With greater amplitude of the inhibitory synapses ($A_i$, Eq. (21)), the stabilisation of the membrane potential is observed only for higher values of $c$ (top row). This is reflected in the firing regularity of the DT-LIF model (bottom row), analogously to the Supplementary Figure 2. The dotted line approximates the separation between the regions where higher $c$ leads to higher / lower membrane potential fluctuations (for the top row) or higher / lower firing regularity (for the bottom row).

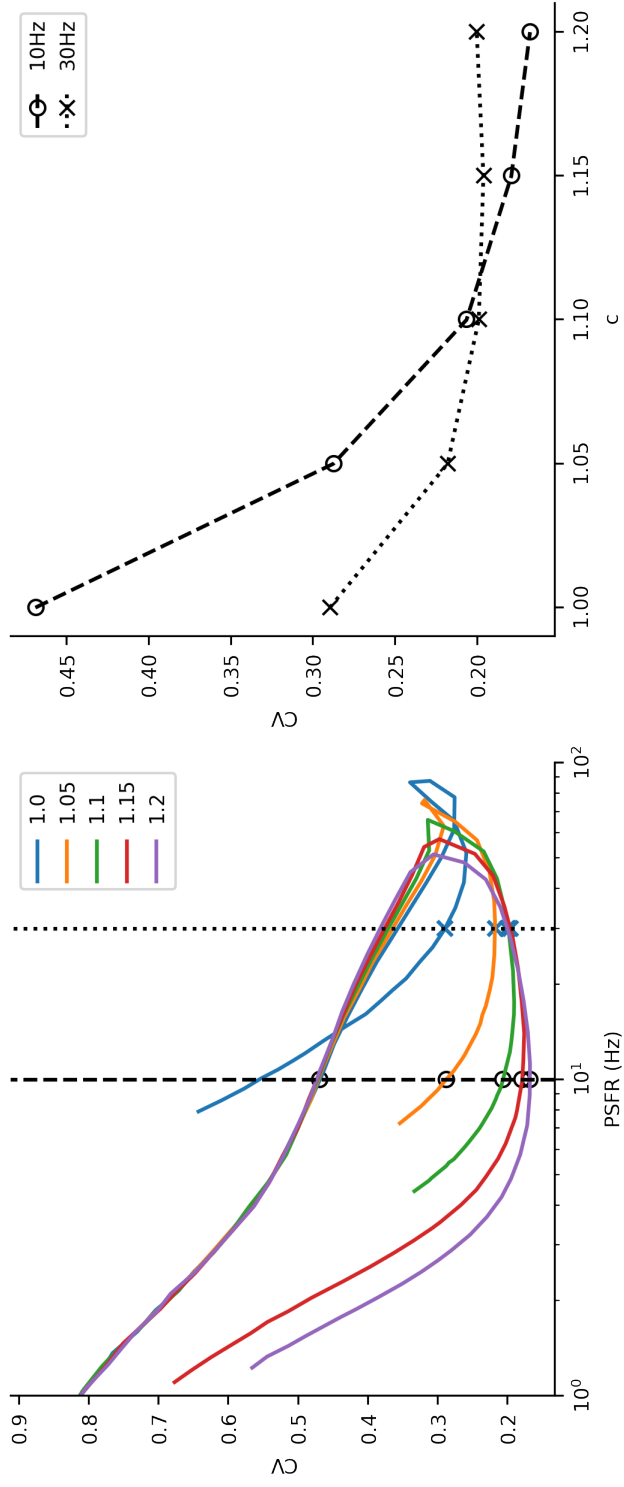

**Supplementary Figure 4: Continuous change of $C_V$ in the HH-DT model.** Left: Same as Fig. 4C, but for $c \in \{1, 1.05, 1.1, 1.15, 1.2\}$. Dashed and dotted vertical lines with points indicate the lowest value of $C_V$ for each $c$. These are then shown in the right panel.


\fi

\end{document}